\newif\ifsubmode
\submodefalse

\newif\ifprintfig
\printfigtrue

\newif\ifemulate
\emulatetrue

\ifsubmode
  \emulatefalse \documentclass[12pt,preprint]{aastex}
  \usepackage{epsfig}
  \received{}
  \accepted{}
  \journalid{}{}
  \articleid{}{}
\else
  \ifemulate
    \documentclass[apj]{emulateapj}
    \received{May 4, 2009}
    \accepted{Dec 8, 2009}
  \else
    \documentclass[preprint]{aastex}
  \fi
  \usepackage{epsfig}
\fi

\shorttitle{Omega Centauri: HST Photometry and Proper Motions}
\shortauthors{Anderson \& van der Marel}

\def\minspt{$\buildrel{\prime}\over .$}

\begin{document}

\title{New Limits on an Intermediate Mass Black Hole\\ 
       in Omega Centauri: I.~Hubble Space Telescope\\ 
       Photometry and Proper Motions\footnote{
               Based on observations with the NASA/ESA 
               {\it Hubble Space Telescope}, obtained at the 
               Space Telescope Science Institute, which is operated by
               AURA, Inc., under NASA contract NAS 5-26555.}} 

\author{Jay Anderson and Roeland P.~van der Marel}
\affil{Space Telescope Science Institute,
       Baltimore, MD 21218, USA; {\tt jayander,marel@stsci.edu}}

\begin{abstract}
We analyze data from the Hubble Space Telescope's Advanced Camera for
Surveys of the globular cluster Omega Cen.  We construct a photometric
catalog of $1.2 \times 10^6$ stars over a $10^{\prime} \times
10^{\prime}$ central field down to below $B_{\rm F435W} = 23$ ($M \sim
0.35 M_{\odot}$).  The 2.5- to 4-year baseline between observations
yields a catalog of some $10^5$ proper motions over a smaller area,
with 53,382 ``high-quality'' measurements in a central $R \lesssim
2$\arcmin\ field.  Artificial-star tests characterize the photometric
incompleteness.  We determine the cluster center to $\sim 1''$
accuracy from star counts using two different methods, one based on
iso-density contours and the other on ``pie-slices''. We independently
confirm the result by determining also the kinematical center of the
HST proper motions, as well as the center of unresolved light seen in
2MASS data. All results agree to within their 1--$2''$ levels of
uncertainty. The proper-motion dispersion of the cluster increases
gradually inwards, but there is no variation in kinematics with
position within the central $\sim 15''$: there is no dispersion cusp
and no stars with unusually high velocities.  We measure for the first
time in any globular cluster the variation in internal kinematics
along the main sequence.  The variation of proper-motion dispersion
with mass shows that the cluster is not yet in equipartition.  There
are no differences in proper-motion kinematics between the different
stellar populations of Omega Cen.  Our results do not confirm the
arguments put forward by Noyola, Gebhardt \& Bergmann to suspect an
intermediate-mass black hole (IMBH) in Omega Cen. They determined
line-of-sight velocity dispersions in two $5'' \times 5''$ fields, and
reported higher motions in their central field.  We find the
proper-motion kinematics to be the same in both fields.  Also, we find
that they (as well as other previous studies) did not accurately
identify the cluster center, so that both of their fields are in fact
$12''$ from the true center.  We also do not confirm the central
density cusp they reported (in part due to the different center, and
in part due to biases induced by their use of unresolved light).  The
surface number-density distribution near the center does not differ
strongly from a single-mass King model, although a shallow cusp may
not be ruled out.  In Paper~II we present new dynamical models for the
high-quality data presented here, with the aim of putting quantitative
constraints on the mass of any possible IMBH.
\end{abstract}

\keywords{catalogs ---
          globular clusters: general ---
          globular clusters: individual (Omega Cen) ---
          stars: kinematics ---
          techniques: image processing, photometric.}


\section{INTRODUCTION}
\label{s.INTRO}

There is much current interest in finding black holes at the centers
of globular clusters (GCs). Supermassive ($\gtrsim 10^6 M_{\odot}$)
black holes are known to exist at the centers of galaxies, and it has
been demonstrated that the black-hole mass is correlated with the
velocity dispersion of the galaxy. If GCs house massive central black
holes that follow the same correlation, then based on their central
velocity dispersions of order $\sim$10 km/s, we would expect to find
an intermediate-mass black hole (IMBH) with a mass of a few thousand
solar masses (Gebhardt, Rich \& Ho 2002). Plausible formation
scenarios exist to form such IMBHs in the centers of globular clusters
from realistic initial conditions (Portegies Zwart \& McMillan
2002). Determining whether globular clusters harbor IMBHs will answer
important questions about how clusters form, and about what kinds of
circumstances give rise to IMBHs (e.g., van der Marel 2004).

An IMBH should induce a power-law cusp in the stellar-density profile
(Baumgardt, Makino \& Hut 2005).  Such cusps are not uncommon in
globular clusters (Noyola \& Gebhardt 2006).  However, various stages
of core collapse can introduce similar cusps, so observing a cusp is
not definitive evidence for an IMBH.  X-ray or radio emission may
point towards accretion onto an IMBH (Pooley \& Rappaport 2006; Kong
2007; Ulvestad, Greene \& Ho 2007), but other explanations for the
emission are difficult to rule out. Moreover, globular clusters
generally have little gas to accrete.  The most unambiguous way to
identify and weigh an IMBH is therefore to find the signature it
induces in the kinematics of nearby stars. There are two ways in which
kinematics could identify an IMBH. First, there may be a general
increase in the velocity dispersion towards the center that cannot be
accounted for by the visible matter. Second, there may be stars moving
faster than would be allowed by the cluster's nominal escape velocity
(Drukier \& Bailyn 2003). One might even hope to observe stars in
Keplerian orbit, as has been seen at the center of our own Galaxy
(Genzel et al.\ 2003; Ghez et al.\ 2005).

Kinematical evidence for IMBHs on the basis of an increase in the
line-of-sight velocity dispersion towards the center has been
presented for the globular clusters M15 and G1 (van der Marel et al.\
2002; Gerssen et al.\ 2002; Gebhardt, Rich, \& Ho 2002,
2005). However, in M15 the implied dark mass could also be attributed
to segregation of dark remnants towards the center (e.g., Baumgardt et
al.\ 2003a), while in G1 the statistical significance of the implied
dark mass is not strong (Baumgardt et al.\ 2003b). The limitations of
these past investigations are due in part to their use of
line-of-sight velocities. Proper motions can generally put more
powerful constraints on an IMBH than can line-of-sight velocities,
because they can be measured for many more stars.  Since proper motions
probe two components of the motion, they also yield better constraints
on the velocity anisotropy and therefore help break the well-known
degeneracy between mass and anisotropy. In cases where one can assume
spherical symmetry, this degeneracy is resolved completely (Leonard \&
Merritt 1989). Furthermore, while spectroscopic line-of-sight velocity
studies are limited to the bright stars, the Hubble Space Telescope
(HST) can observe proper motions for the much more plentiful
main-sequence stars. This gives better statistics to probe closer to
the center. IMBH limits from HST proper motions have been presented
for M15 (McNamara et al.\ 2003 and van den Bosch et al. 2006)
and 47 Tuc (Mcaughlin et al.\ 2006). These studies relied, at least in
part, on data from the Wide Field and Planetary Camera 2 (WFPC2)
instrument. Several studies have since been started or planned to
study this problem exclusively with data from newer Advanced Camera
for Surveys (ACS) or Wide Field Camera 3 (WFC3) instruments for larger
samples of clusters (e.g., GO-9835 and GO-10474, PI-Drukier;
GO-10401/10841, PI-Chandar; GTO-10335/11801, PI-Ford; and GO-11609,
PI-Chaname). Two epochs are already in-hand for several of the target
clusters in these samples and results may start to come out soon.

Recently, Noyola, Gebhardt, and Bergmann (2008, NGB08) presented a new
study of the globular cluster Omega Cen. They used Gemini IFU
spectroscopy to measure the line-of-sight velocity dispersion of
unresolved light in two $5'' \times 5''$ fields, one at the cluster
center and one at $R=14''$ from the center. The dispersion in the
central field ($23.0 \pm 2.0$ km/s) exceeded that in the off-center
field ($18.6 \pm 1.6$ km/s). Based on this increase towards the center
they argued for the presence of an IMBH of mass $4.0_{-1.0}^{+0.75}
\times 10^4 M_{\odot}$. NGB08 also measured the surface-brightness
profile of unresolved light from HST/ACS images. They found it to have
a shallow central cusp of logarithmic slope $\gamma = 0.08 \pm 0.03$,
also consistent with the presence of an IMBH.

A considerable amount of ACS data of Omega Cen already exists in the
HST Data Archive. In the present paper we collect and analyze these
data to obtain photometric and proper-motion catalogs for very large
numbers of stars (of order $10^6$ photometric measurements and $10^5$
proper motions, respectively). We use these catalogs to perform
detailed star-count and kinematical analyses. The present paper is
confined to the observational domain as much as possible, and we focus
on testing and augmenting the results of NGB08. Detailed dynamical
modeling of the new data is presented in a companion paper (van der
Marel \& Anderson 2009; Paper~II).

This paper is organized as follows. In Section~\ref{S.obs} we give an
overview of the three ACS/WFC data sets that are part of our study.
In Section~\ref{S.redux} we construct a photometric catalog covering
the inner part of the cluster, and we determine proper motions for the
stars that could be measured well in two epochs. In
Section~\ref{S.center} we determine the position of the cluster center
using our new star lists and proper motions, as well as ground-based
2MASS data. In Section~\ref{S.sdp} we derive the number-density
profile. In Section~\ref{S.pmanal} we do an initial analysis of the
proper-motion kinematics (with more detailed analysis following in
Paper~II), and we calculate the proper-motion dispersions in the
fields observed by NGB08. Our results for the cluster center, density
profile, and kinematical gradient are all different from those
presented by NGB08.  We do not confirm the arguments put forward by
them to suspect an IMBH in Omega Cen. In Section~\ref{S.Exploits} we
combine our proper motions and photometry to determine whether the
cluster core is in equipartition and to determine whether the
different stellar populations in the cluster have different kinematics.
Finally, in Section~\ref{S.concl} we present the conclusions of our study.


\section{OBSERVATIONS}
\label{S.obs}
This paper draws on three HST data sets, listed in Table~\ref{tab01}. 
All three were taken with the Wide-Field Channel (WFC) of the 
Advanced Camera for Surveys (ACS), which is made up of two 
2048$\times$4096-pixel detectors with a pixel scale of about 
50 mas/pixel, covering a rhombus-shaped field that is roughly 
3\minspt4 on a side.


\begin{table}
\begin{center}
\caption{Data available for the proper-motion study. \bigskip}
\begin{tabular}{|c|c|c|}
\hline 
 Dataset  &  CENTRAL FIELD      & OUTER FIELD \\
\hline 
GO-9442   & 3$\times$340s F435W & 3$\times$340s F435W \\
(2002.49) & 3$\times$340s F625W & 3$\times$340s F625W \\
          & 4$\times$440s F658N & 4$\times$440s F658N \\
          & (Pointing 5)        & (Pointings 1 and 4) \\
\hline 
GO-10252  &  ---                & 5$\times$340s F606W \\
(2004.95) &  ---                & 5$\times$340s F658N \\
\hline 
GO-10755  & 4$\times$80s F606W  & --- \\
(2006.56) & 4$\times$90s F814W  & --- \\
\hline 
\end{tabular}
\label{tab01}
\end{center}
\end{table}

The first data set is GO-9442 (PI-Cool).  It was taken over the period 
between 27 and 30 Jun 2002 and consists of a 3$\times$3 mosaic of 
pointings, centered on the cluster center.  Each pointing has 3 deep 
exposures and 1 shallow exposure in each of F435W ($B_{\rm F435W}$) and 
F625W ($R_{\rm F625W}$) and 4 deep observations through F658N (H-alpha, 
$H_{\rm F658N}$).  Each of the deep observations was offset relative 
to the others, so that no star would fall in the 50-pixel-tall 
inter-chip gap in more than one deep exposure for each filter.  The 
mosaic of pointings covers the inner $10^{\prime}\!\times\!10^{\prime}$, 
extending out to about two core radii ($r_c \sim$ 2\minspt5).  The 
left panel of Figure~\ref{fig01} shows the field coverage for the 
$B_{\rm F435W}$ filter.  HST mis-pointed in the last pointing (upper 
right), and there is a small gap in the field coverage there.  We were
careful to include the effect of this gap in our star-count analysis.

\ifemulate
\begin{figure}[t]
\epsfxsize=0.99\hsize
\centerline{\epsfbox{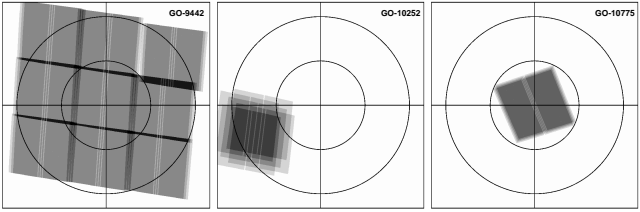}}
\figcaption{At each point in the field centered on Omega Cen, we show the 
         number of exposures available for each of the two broadband colors 
         for the three data sets.  The circles indicate one core radius 
         (2\minspt5) and two core radii.  See text for a description of
         the data in each program.  The baseline between the first and
         second panel is 2.46 years, and between the first and third 
         is 4.07 years.
         \label{fig01}}
\end{figure}
\fi

The second data set is GO-10252 (PI-Anderson).  This program was taken on 
11 Dec 2004, and consists of 5 deep exposures and one shallow exposure in 
each of F606W $(V_{\rm F606W})$ and F814W ($I_{\rm F814W})$.  The goal of 
these observations was to construct accurate PSF models and an accurate 
distortion solution for the two most commonly used filters, thus the 
various exposures have large offsets of about 500 pixels between them.  
Unfortunately, at the time
these observations were taken, the camera was considerably out of 
focus (it was adjusted soon afterwards), and the images were therefore
not useful for their original aim of probing the ``typical'' PSF 
behavior and distortion solution.  However, they are still useful 
for our purposes here, since we can tailor-make PSFs for these images,
and can measure our motions locally, so that small errors in the 
global distortion solution will not impact our results.  The middle 
panel of Figure~\ref{fig01} shows that this field is centered near 
the south-east corner of the mosaic, along the cluster's major axis.  
The baseline between this and the first dataset is 2.49 years.

The third data set is GO-10775 (PI-Sarajedini).  These observations were 
taken as a part of a large survey of 65 globular clusters, and consist
of 4 deep exposures and one shallow exposure in each of F606W and F814W.
The observations were taken on 22 Jun 2006, so the baseline between this
set and the first is 4.07 years.  The right panel of Figure~\ref{fig01} 
shows that this field is centered on the cluster, and on the GO-9442 mosaic.

In the following section, we describe how we constructed an extensive 
catalog based on the $B_{\rm F435W}$ and $R_{\rm F625W}$ exposures of the 
$10^{\prime}\times10^{\prime}$ GO-9442 data set; then we 
cross-identify the same stars in the other data sets and measure 
proper motions.


\section{REDUCTIONS}
\label{S.redux}
In this section we describe how we constructed a reference frame 
and a catalog of photometry and proper motions.  In the subsequent 
sections, we will then use this catalog to determine the cluster 
center and study the surface-density and velocity-dispersion profiles.

\subsection{Reference frame}
\label{SS.redux_refframe}
The first task in the analysis is to construct a reference frame.  We 
began by measuring all the bright, isolated, unsaturated stars in all 
the deep exposures of GO-9442.  There are 36 exposures total in H-alpha, 
and 27 in each of $B_{\rm F435W}$ and $R_{\rm F425W}$.  To do this, we used the 
publicly available software program {\tt img2xym\_WFC.09x10} 
(Anderson \& King 2006), which employs an empirically constructed, 
spatially variable library PSF to determine a position and a flux 
for each star in each exposure.  The positions are corrected for 
distortion using the corrections in Anderson (2005).  

We began by analyzing the nine major pointings independently.  For
each, we took the central $R_{\rm F625W}$ observation as the basis for the
reference frame of that pointing, linearly transformed the star
positions from all the other images of that pointing into that frame,
and determined an average position for each star that was found in at
least three exposures.

The next step was to stitch together the nine overlapping frames.  There 
was only minimal overlap between the frames.  We did not want to allow for 
linear terms in the frame-to-frame transformation, since small errors in 
such terms could lead to large errors in the extrapolated
field, so we
stitched the pointings together by solving only for offsets and rotations, 
assuming the frames to have the same scales and off-axis linear terms.  
We first added the directly adjacent side fields to the central frame,
than added the corner fields to this plus-shaped intermediate frame.

To improve this reference frame, we found a linear transformation from
each exposure into the new frame, based on the positions of common stars.  
For each star in the reference frame, we thus had between 7 and 40 estimates 
for its position (depending on how many images overlapped at that point),
and we averaged these positions together to improve the reference frame.
After a few such iterations, we found the remaining stitching errors to 
be less than 0.01 pixel. 

The above reference frame was constructed in a distortion-corrected 
frame based on the central pointing.  To align our frame with RA and Dec, 
we found the linear transformation from this frame into the frame constructed 
in Anderson et al.\ (2008; hereafter A08), based on the GO-10775 Treasury 
data, which has a pixel scale of 50 mas/pixel and was aligned with RA and 
Dec, and has the Harris (1996) cluster center at [3000,3000].  
(We will refer to this as the ``traditional'' center, since it is based
on the literature at the time.)  We conformally transformed our reference 
frame into this frame and added 4000 to each coordinate, so that our 
entire field would be contained within the range [1:14000,1:14000], 
with the center close to [7000,7000].  We refer to this frame as the 
``master'' frame.

At this point we made a digression, to put our reference frame on an
absolute basis, even though it is not needed in the present study.
The guide-star catalog, and hence the header information, can contain 
astrometric errors of up to 1.5 arcsecond (Koekemoer et al.\ 2005), so we 
cross-identified over 6000 of our stars with the 2MASS catalog in order 
to construct a more accurate absolute astrometric calibration of our 
reference frame.  We found that the 2MASS frame was offset by 0.5 arcsecond 
in declination and 2 arcseconds in right ascension from the frame given 
in the GO-10775 {\tt drz}-frame header.  We also found that there was 
a small orientation difference ($0.1^{\circ}$) between the {\tt drz}-based 
frame and the 2MASS frame.  The WCS (world-coordinate system) parameters 
in the {\tt fits} headers of the stacked images that we provide were 
constructed to match as well as possible the positions of stars in our 
frame with those in the 2MASS catalog.  Skrutskie et al.\ (2006) argue 
that the 2MASS positions for well-measured stars should be good to about 
50 mas in a random sense, and good to 15 mas in a systematic sense. 

We emphasize that our proper motions do not depend greatly on the quality
of our astrometric solution.  Each star will be measured relative to 
its nearby neighbors using only individual flat-fielded ({\tt \_flt}) 
images; the reference frame contributes only to our scale and 
orientation.  Nonetheless, for future use of the catalogs provided here
it is still desirable to have the best possible reference frame.

\subsection{Image stack}
\label{SS.redux_stack}
The transformations from the frame of each exposure into the master
frame allow us to generate stacked images of the field in this frame.
Our images were generated in a manner akin to {\tt drizzle} with a
{\tt pixfrac} parameter of 0 (see Fruchter \& Hook 2002).  These
stacked images provide a simple representation of the field that
allows us to independently evaluate how our star-finding algorithms
performed (our finding was done on the individual images, not on the
stacks).

The stacked images are 14000$\times$14000 pixels, covering 11.6 arcminutes
on a side, and were constructed for the F435W, F625W, and F658N filters
of GO-9442.  We provide these images as a part of the data release with 
this paper.

\subsection{Constructing the Catalog}
\label{SS.redux_catalog}
Once we had determined a reference frame, we could construct a star
catalog in this frame.  To do this, we used the same software program
that was used for the reductions for the 65 clusters in GO-10775.  The
details of this program can be found in A08.  Briefly, the routine
reads in the available images (shallow and deep) for two different
filters ($B_{\rm F435W}$ and $R_{\rm F625W}$, in this case) and auxiliary
information that allows it to map each exposure into the reference
frame, both astrometrically and photometrically.  It then goes through
the reference frame in patches that are 25$\times$25 pixels in size,
and identifies all the stars in each patch, measures them, and records
their parameters in a file.  The routine is able to deal well with
short and deep exposures; if a star is saturated in the deep
exposures, it is found in the short exposures where possible.  It also
finds stars iteratively, first finding the bright stars, then removing
them to find the fainter stars that might not stand out distinctly in
the un-subtracted images.  Finally, the routine is robust against
identifying PSF artifacts or diffraction spikes as stars.

The GO-9442 data set differs from the typical GO-10775 cluster data set.
Here, the master frame was 14000$\times$14000 WFC pixels in size, and 
consisted of a mosaic of pointings, whereas in the GO-10775 set, the 
observations consisted of medium-sized dithers about one central pointing 
and could fit comfortably within a 6000$\times$6000-pixel region.  The
typical point in the field under study here had coverage of 3 $B$ and 
3 $R$ observations.  In order to be included in the catalog, we insisted 
that a star be found in at least 4 out of these 6 observations independently; 
if there were fewer exposures available at a given point in the field,
the criteria were relaxed accordingly (see A08).

The automated finding program identified 1,164,317 stars in the 
$B_{\rm F435W}$ and $R_{\rm F625W}$ images.  We plotted the star lists on 
top of the stacked images to verify that the routine had identified all 
the stars we expected it to.  This inspection confirmed that there were 
very few stars that were missed by the automated routine.  The few missed 
stars fall into two classes:  medium-brightness stars in the vicinity of 
saturated stars, and blended pairs of stars that had one star that was
brighter in $B_{\rm F435W}$ and the other brighter in $R_{\rm F625W}$.   
The former were passed over by the program because they were too likely 
to be artifacts near saturated stars, and it was better to err on the 
side of caution.  The latter were missed because we required each star
to be found as a dominant peak in at least four independent images.  
Both of these loss mechanisms can be quantified with artificial-star 
tests (see below).

To get an idea of how many real stars were missed, we ran the same
finding program on the F658N images (which had less saturation and
four exposures at each pointing), requiring a star to be found in at
least 3 out of 4 exposures, and cross-identified the stars in the two
catalogs.  There were about 30,000 stars (3\% of the total) that were
found in the H-alpha images that were not found in the $B_{\rm F435W}$ and 
$R_{\rm F625W}$ observations.  We looked at where these missing objects were
located, and they were almost exclusively found in the situations
detailed above.

In Figure~\ref{fig02}, we show the central region of the cluster.
The stars found jointly
in $B_{\rm F435W}$ and $R_{\rm F625W}$ are marked in 
yellow.  The stars that were found in $H_{\rm F658N}$ (H-alpha) but 
not in $B_{\rm F435W}$ and $R_{\rm F435W}$ are marked in red.  Of 
course it would be good to have a list that contains every single 
star in the cluster, but a compromise must always be made between 
including the marginal star and including image artifacts as stars.  
There will always be stars that are missed; the important thing is 
to be able to quantify this incompleteness.

\ifemulate
\begin{figure}[t]
\epsfxsize=0.99\hsize
\centerline{\epsfbox{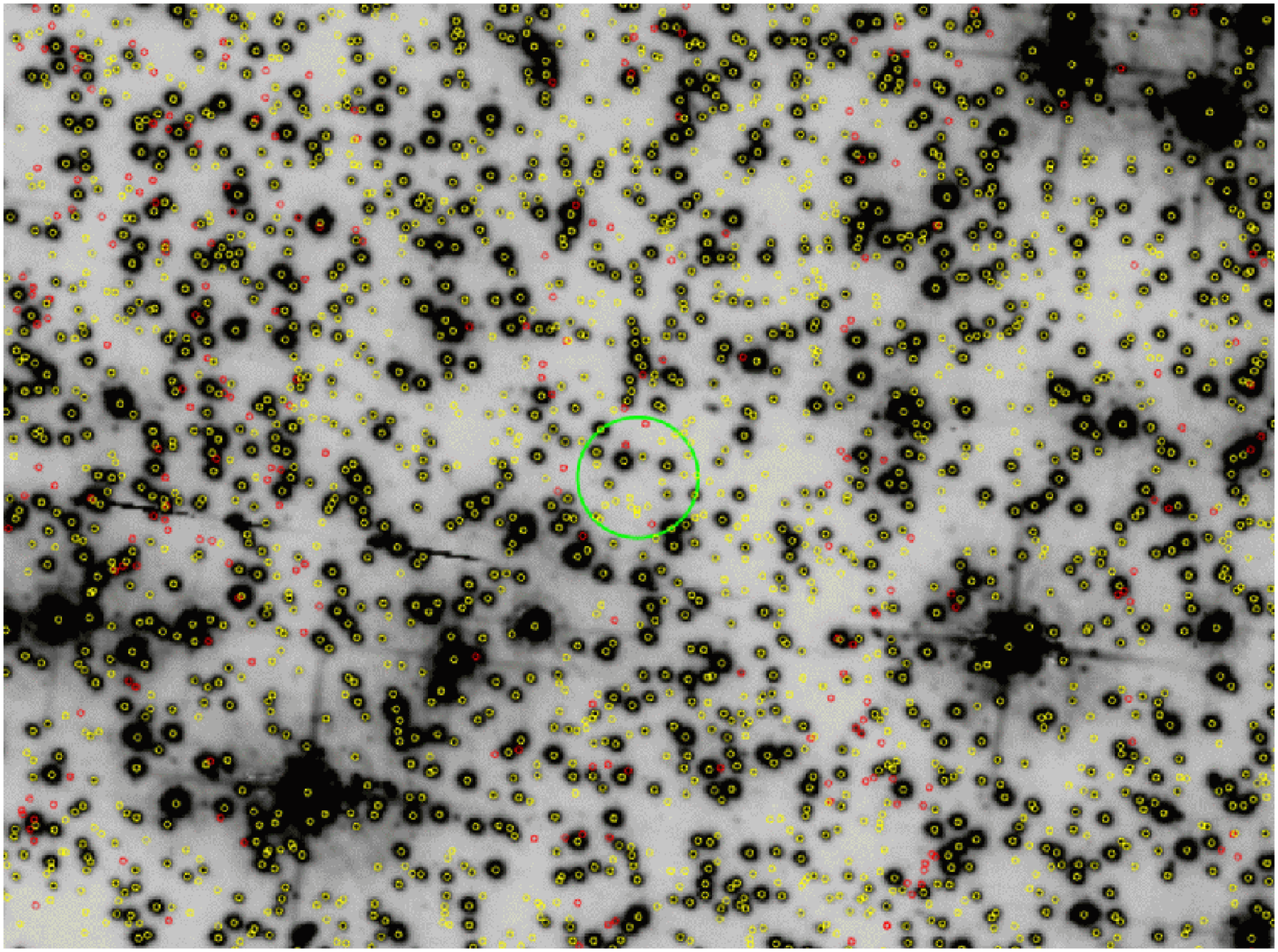}}
\figcaption{A stack of the deep F625W images at the center of the cluster.
         The yellow circles identify the stars found in the 
         $B_{435}+R_{625}$ list.  The red-circled stars were found in the 
         $H_{658}$ list, but not the $B_{435}+R_{625}$ list.  The 
         large green circle is the location we have identified for 
         the center (see Section~\ref{S.center}).  The field shown 
         is 21\arcsec $\times$ 16\arcsec . 
         \label{fig02}}
\end{figure}
\fi

\subsection{Calibrating the photometry}
\label{SS.redux_calib}
The fluxes for the stars measured in the above routine are reported in 
instrumental magnitudes, $-2.5\,{\rm log_{10}}Z$, where $Z$ is the scaling 
that matches the effective PSF model to the stellar profile.  The PSF model 
was normalized to have a flux of 1.00 within a radius of ten {\tt \_flt} 
pixels.  The short exposures have been zero-pointed to match up with the 
deep exposures, therefore the task of calibration simply involves 
determining the zero-point that will bring the instrumental photometry
of the deep exposures into the VEGAMAG system.

Since the GO-9442 data set has already been calibrated by Villanova 
et al.\ (2007), we simply adopted the photometric zero points from that 
project.  They are $ZP_{\rm F435W} = 32.043$, $ZP_{\rm F625W} = 31.991$, 
and $ZP_{\rm F658N} = 28.904$.  Figure~\ref{fig03} shows our 
$(B_{\rm F435W}\!-\!R_{\rm F625W})$ color-magnitude diagram (CMD) in 
both the instrumental and in the calibrated systems.  We will continue
to report many of the analyses here in the instrumental system, since 
it is easier to assess errors in terms of signal-to-noise ratio when 
working directly in photo-electrons.  For reference, saturation in the
deep images sets in at an instrumental magnitude of around $-13.75$, and 
a star with a magnitude of $-10.00$ has a signal-to-noise  ratio
of 100.  In the rest of the paper, when referring to the instrumental 
system, we will report magnitudes as $m_{\rm F435W}$, and when referring 
to the calibrated VEGAMAG photometry, we will refer to $B_{\rm F435W}$.

\ifemulate
\begin{figure}[t]
\epsfxsize=0.99\hsize
\centerline{\epsfbox{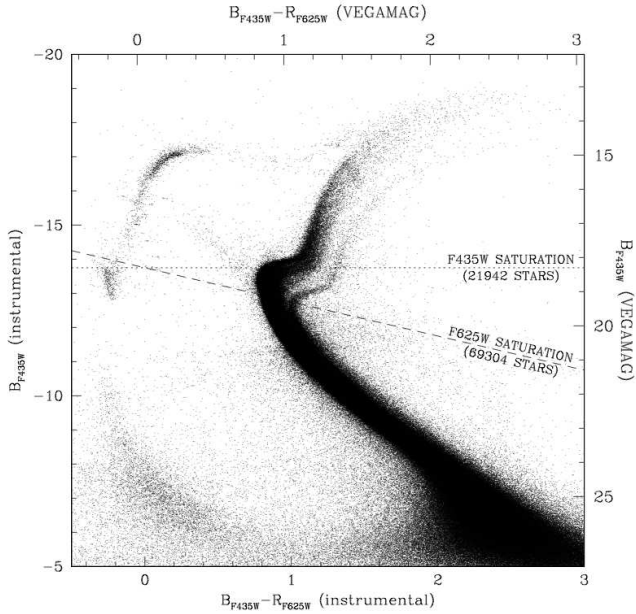}}
\figcaption{The CMD of all the stars found in the automated search of the 
         F435W and F625W exposures of GO-9442.  The level where saturation
         sets in for the deep exposures is indicated, as well as the number of
         stars that are saturated.  Most of the stars above the deep-image
         saturation line were measured well in the short exposures.  
         Saturation in the short exposures sets in at about 4 magnitudes 
         brighter than this.  These particularly bright stars were measured 
         by fitting the PSF to the unsaturated part of the star's profile.
         Note the multiple main-sequence turn-offs, which have been 
         well-discussed in the literature (i.e., Ferraro et al.\ 2004 and 
         Villanova et al.\ 2007).
         \label{fig03}} 
\end{figure}
\fi

\subsection{Artificial-star tests}
\label{SS.redux_AS}
The GO-9442 images are not terribly crowded.  There are about 1.2 million
stars in 14K$\times$14K pixels, thus the typical separation is about
10 pixels.  Nevertheless, the fact that there are quite a few
extremely bright stars makes it impossible to find all the faint
stars.  And even though the density changes by less than a factor of
four from the center to the corners of the field, it is still
important to have an idea of the completeness for each brightness
level of star at different places within the field.  Therefore, we
decided to run artificial-star tests in order to gauge the
incompleteness and measurement quality.

The mechanism for artificial-star (AS) tests is described thoroughly in 
A08.  Briefly, since our finding software operates on one small patch 
of the field at a time, we can afford to do artificial-star tests in 
serial, one at a time, rather than doing them in parallel in many batches.
One great benefit of this is that artificial stars never interfere
with each other, so by throwing many stars in successively we can simulate
throwing in a high density of them.  This makes it easy to do a 
detailed study of the completeness in the vicinity of the center.

We performed 500,000 individual artificial-star tests.  For each, we
chose a random F435W instrumental magnitude between $-14$ (saturation)
and $-5$, and chose the F625W magnitude that placed the star along the
fiducial main sequence.  The artificial stars were inserted with a
flat distribution in radius, so that we would probe the central
regions more than the outer regions.  For example, we ended up
inserting over 700 stars within the central arcsecond in radius, and
over 51,000 stars within the central 10 arcseconds.

As a demonstration of how the AS tests worked, in Figure~\ref{fig04}
we show the same central field as in Figure~\ref{fig02}, and the
locations where the artificial stars were and were not recovered.  In
this region, we observed  261 real stars between $-10.5$ and $-11.5$ in 
F435W instrumental magnitudes.  We ran 3847 artificial-star tests in this 
magnitude range over this 225-square-arcsecond region, and recovered 3402
of them, resulting in a completeness of 88 \%.

\ifemulate
\begin{figure}[t]
\epsfxsize=0.99\hsize
\centerline{\epsfbox{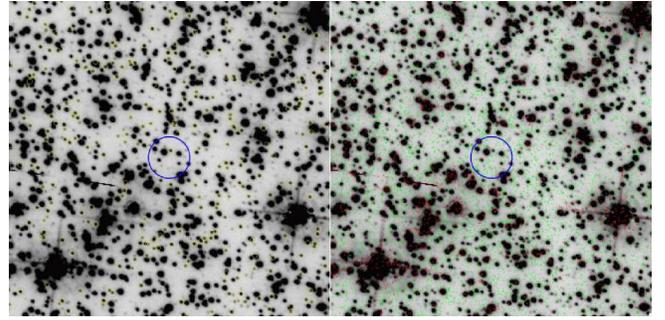}}
\figcaption{Central swath of the stacked image (15\arcsec\ by 15\arcsec\
         in each panel).  On the left, the yellow circles indicate 
         stars that were found between $-10.5$ and $-11.5$ (in 
         F435W instrumental magnitudes).  On the right we 
         show where stars in the same brightness range were successfully 
         recovered from our artificial-star tests (green), and where 
         they were not recovered (red).  The center (determined in 
         Section~\ref{S.center}) is shown in both panels with a 
         blue 1\arcsec -radius circle around it.  
         \label{fig04}} 
\end{figure}
\fi

Figure~\ref{fig05} shows the results of the artificial-star tests for 
different radial bins as a function of instrumental magnitude.  It is 
clear that the completeness at the center is well over 90\% for the 
brightest stars, and is over 75\% throughout the cluster for stars with 
instrumental magnitudes brighter than $-10$, which corresponds to stars 
about 4 magnitudes below the upper SGB. 

\ifemulate
\begin{figure}[t]
\epsfxsize=0.99\hsize
\centerline{\epsfbox{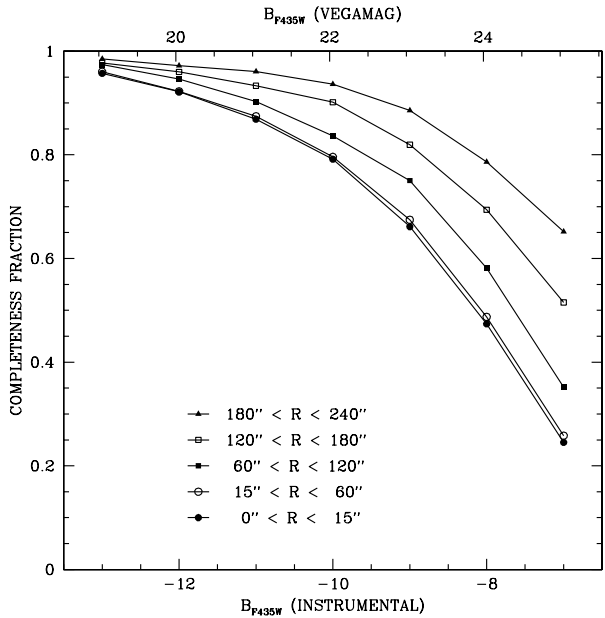}}
\figcaption{The completeness as a function of magnitude for stars at different
         distances from the cluster center.
         \label{fig05}}
\end{figure}
\fi
 
\subsection{Measuring the Proper Motions}
\label{SS.redux_PMs}
Figure~\ref{fig01} shows that proper motions can be measured in two fields:
one field that is largely centered on the cluster and another that is east 
south-east of the center by about 4 arcminutes.  Table~\ref{tab01} lists 
the observations we have available for each field.  

Our procedure for measuring proper motions was to measure the stars 
individually in each exposure for each epoch, and then to combine the
many independent measurements to construct proper motions for the stars
that can be found in multiple epochs.  The agreement among the many 
independent observations for each star at each epoch gives us a handle 
on the errors in the positions, and hence in the displacements and proper 
motions.  

The GO-10252 images are well matched depth-wise to the GO-9442 images, 
thus we should expect the same stars to be measured equally well in both 
epochs.  By contrast, the GO-10775 images, which cover the central field, 
are a factor of four shallower than the GO-9442 data set, and therefore have 
different signal-to-noise ratios for the same stars.

\subsubsection{Measuring the star positions}
\label{SSS.redux_PMs_pos}
The first step in determining proper motions is to measure an 
accurate position for each star in each individual exposure at each 
epoch.  To do this, we once again ran the star-measuring program 
{\tt img2xym\_WFC.09x10} to construct a list of sources in each 
exposure, using empirically determined PSFs with a spatially-constant 
perturbation to account for breathing-related focus changes from 
exposure to exposure.  

The routine produced a list of positions and instrumental magnitudes
for the reasonably bright and 
isolated stars identified in each 
exposure.  Since our aim was to include only the unsaturated stars 
that could be measured accurately, we adopted finding parameters 
such that we would find unsaturated stars with at least 250 counts 
in their brightest four pixels and with no brighter neighbors within 
4 pixels.  We also confined our attention to the deep exposures 
for each filter/epoch combination, since there is generally only one 
short exposure at each location and it is hard to get a good handle 
on astrometric errors without multiple exposures.  Finally, we treated
the central and adjacent major-axis proper-motion fields separately,
since there is only marginal overlap between them.  

\subsubsection{The central field}
\label{SSS.redux_PMs_cen}
For the central field, we have 10 exposures in the first epoch in filters 
$B_{\rm F435W}$, $R_{\rm F625W}$, and $H_{\rm F658W}$, and 8 exposures in 
the second epoch in filters $V_{\rm F606W}$ and $I_{\rm F814W}$.  We 
reduced each exposure as described above, producing a list of $x$, $y$, 
and $m$ (instrumental magnitude) for each source.  We treated the two 
WFC chips for each exposure independently, so we had a total of 36 lists of 
stars, one for each chip in each exposure.  We corrected the positions
for distortion and cross-identified each list with the reference-frame
list, allowing us to develop a least-squares linear transformation from 
each exposure into the reference frame.  Since the central field covers 
only a fraction of the GO-9442 field, we restricted ourselves here the 
subset of 422,561 stars contained within the region [4000:9999,4000:9999].

The above linear transformations yielded an estimate for the
reference-frame position for each star from each exposure in which it
was found.  These global transformations can contain some small
systematic residuals, due to variations in the distortion solution
caused by breathing or other effects.  The consequence of these errors
is that the transformed positions for the stars in one part of the
reference frame are all shifted by some (small) amount.  In order to
remove these residual transformation artifacts from the data, we
determined a local adjustment for each observation of each star as
follows.  We first identified all the neighboring stars within a
100-pixel radius that were unsaturated and had a S/N $>$ 40 (excluding 
the target star itself).  For each exposure where the star was observed, 
we then found a robust average offset between the globally transformed 
positions of the neighbor stars for that exposure and the average 
reference-frame positions for the stars.  This average offset provided 
the correction to the transformed position of the target star for that 
exposure.

Figure~\ref{fig06} shows the residuals before and after this
correction for the first image for each epoch.  The typical adjustment
is 0.01 pixel, which is about the accuracy that we expect for the
static distortion solution (Anderson 2003).  But not all of this
adjustment comes from static distortion-solution error; much of it
clearly varies from exposure to exposure due to breathing-related
changes in focus.  The effect of this correction (and, indeed, of the
star-based transformations themselves) is that the motion of each star
is measured relative to that of the average motion of its neighbors.  
Since outliers are rejected in the process of determining the average 
motion of the neighbors, this means that each star's motion is computed 
relative to the bulk cluster motion at its location.  
Section~\ref{SSS.redux_PMs_about} discusses what this calibration 
procedure means in practice for the motions that we measure.

\ifemulate
\begin{figure}[t]
\epsfxsize=0.99\hsize
\centerline{\epsfbox{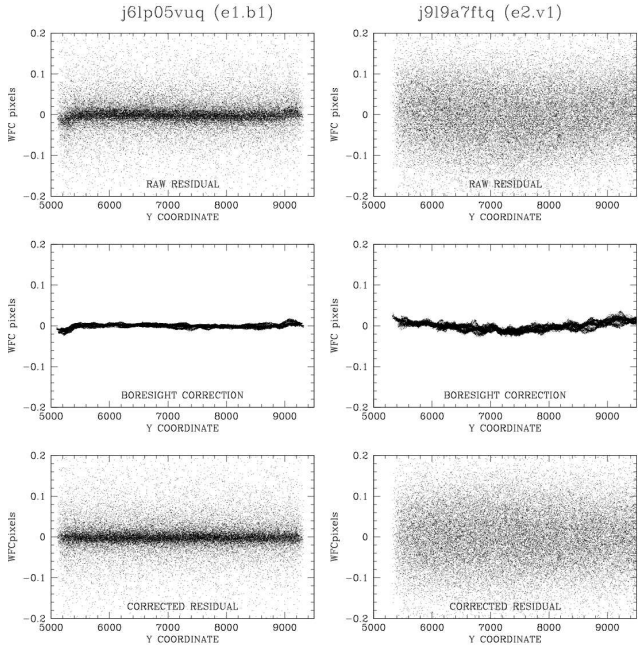}}
\figcaption{(Left) In the top panel, we show the $x$-residual between the 
         first F435W image in GO-9442 ({\tt j6lp05vuq}) and the reference 
         frame for a 100-pixel-tall strip through the cluster center.
         Since the reference frame is based on images taken at that
         epoch, these residuals should show just distortion errors and 
         should have no proper motions.  The middle panel shows the 
         local ``boresight'' correction used for each star.  The bottom 
         panel shows the residuals after the local adjustment.  
         (Right) Same, but for the first exposure in the second epoch 
                 ({\tt j9l9a7ftq}).  These residuals contain both 
                 distortion errors {\it and} actual proper motions, 
                 and therefore show more scatter than in the left panels.
         \label{fig06}}
\end{figure}
\fi

At this point, we had in hand corrected reference-frame positions 
for a large number of stars measured in a large number of independent 
exposures.  We combined the data in several ways.  First, for each 
filter/epoch combination, we averaged all the data together to arrive 
at an average position, magnitude, and RMS error for each star.  We 
also combined the astrometric data together by epoch.  (Since the 
F658N data have a lower signal-to-noise ratio at the faint end, we 
include F658N astrometry only when the star had an instrumental 
magnitude of $H_{\rm F658N} < -10$, indicating a S/N of 100). 

The proper-motion results for the central field are illustrated in 
Figure~\ref{fig07}.  Each data set consisted of 3 to 4 observations 
of each star.  We constructed an average reference-frame position for
each star in each data set, then computed the error in this position
from the RMS about this average.  The two-dimensional error in the 
average position at a single epoch is typically 0.01 pixel.  This error 
was achieved for the data set for each filter at each epoch.  

\ifemulate
\begin{figure}[t]
\epsfxsize=0.99\hsize
\centerline{\epsfbox{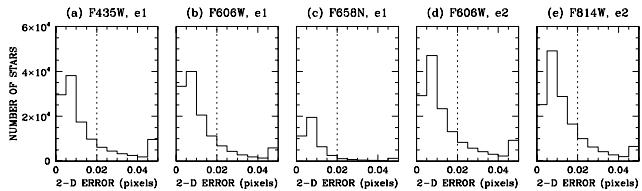}}
\figcaption{Panels (a) through (c) show the distribution of
         $\sqrt{\sigma_x^2+\sigma_y^2}$, the total two-dimensional error in 
         the position for the first-epoch F435W, F625W, and F658N data,
         respectively. This error was computed by taking the RMS of the
         scatter among the multiple observations of the master-frame 
         position for each star and then dividing by the square root 
         of the number of observations.  The dotted line drawn at 
         0.02 pixel indicates the boundary between the well and poorly
         measured stars.  Panels (d) and (e) show, respectively, the 
         same for the F606W and F814W data of the second epoch.
         \label{fig07}}
\end{figure}
\fi

The proper motions were constructed by subtracting the first-epoch
positions from the second-epoch positions and dividing by the time
baseline.  The multiple data sets we had at each epoch allowed us to
construct two independent observations of the proper motions.  We
constructed one displacement comparing the second-epoch F606W positions
with the first-epoch F435W positions, and a second displacement by 
comparing the second-epoch F814W with the first-epoch F625W.  The 
top panels of Figure~\ref{fig08} compare these two independent 
measurements.  The distribution along the 45$^\circ$-direction 
represents the actual motion measured, found to be the same in both 
halves.  The distribution in the 135$^\circ$-direction represents 
the error in our measurements.  The typical displacement between epochs 
is about 0.15 pixel.  The displacement of the typical star is measured 
with a fractional error of less than 10\% (from the aspect ratio of 
distribution in the upper panels).

\ifemulate
\begin{figure}[t]
\epsfxsize=0.99\hsize
\centerline{\epsfbox{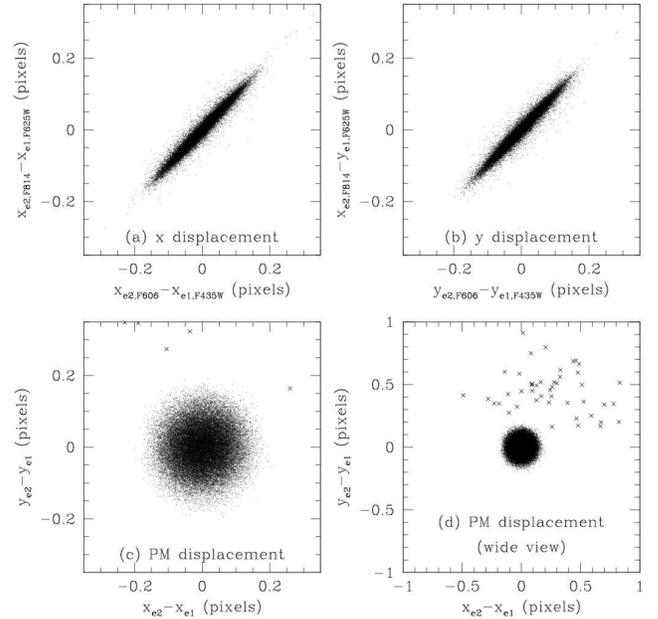}}
\figcaption{Panel (a) compares two independent measurements of the $x$ 
         displacement between epoch-2 and epoch-1.  The agreement between 
         the two is an indication of the quality of the proper motions.  
         Panel (b) shows the same for the $y$ displacement
         Panel (c) shows the resulting 2-d displacement distribution, after
         combining the data from all filters for each epoch.
         Panel (d) zooms out to show the cluster and field stars (points 
         beyond 0.25 pixel are displayed as crosses, for clarity).  The time
         baseline between these epochs is 4.07 years.  The pixel scale
         is 50 mas/pixel, and x-axis of the reference frame is aligned with 
         the negative right-ascension axis.
         \label{fig08}}
\end{figure}
\fi
 
We combined all the data for each epoch and constructed a single 
proper motion for each star, which we plot in the lower panels as 
vector-point diagrams.  There are very few outlier stars, meaning that
we have done a good job selecting stars with small errors.  It is 
worth pointing out that our finding criteria have not selected 
against high-motion stars:  the field stars in this diagram demonstrate 
that we are finding stars with displacements of up to 1 pixel.  Thus, 
we are clearly sensitive to cluster members with more than 10 times the 
typical motion.  The baseline for this central field is 
4.07 years, thus a displacement of 0.08 pixel corresponds to about 
0.02 pixel per year in each coordinate, or 1 mas/yr.

We measured proper motions for the 108,507 stars that could be found 
in at least two images of each epoch, with an RMS of position measurements 
within each epoch of less than 0.03 pixel (per coordinate).  In addition, 
we defined a ``high-quality subset'' of the data that should have more
uniform PM errors by selecting those stars for which this position RMS
was less than 0.02 pixel and for which the instrumental $m_{\rm F435W}$ 
photometry was brighter than $-11$.  Figure~\ref{fig09} shows a CMD of
the stars that qualify for the high-quality subset.  Our high-quality 
motions are limited to stars on or below the sub-giant branch (SGB) and 
those brighter than $m_{\rm F435W} = -11$, a few magnitudes below the 
turnoff.  Stars much brighter than the SGB were saturated in the deep 
GO-10775 exposures.  Most of the proper-motion analysis in this paper 
and Paper II will use this high-quality sample of  53,382 stars.  
Within this sample, the typical proper-motion error is better than 
0.1 mas/yr in each coordinate, corresponding to 0.008 pixel over the
4-year baseline.  Over the magnitude range where we could measure 
reliable motions, we measured between 80\% and 90\% of {\it all} 
the stars.

\ifemulate
\begin{figure}[t]
\epsfxsize=0.99\hsize
\centerline{\epsfbox{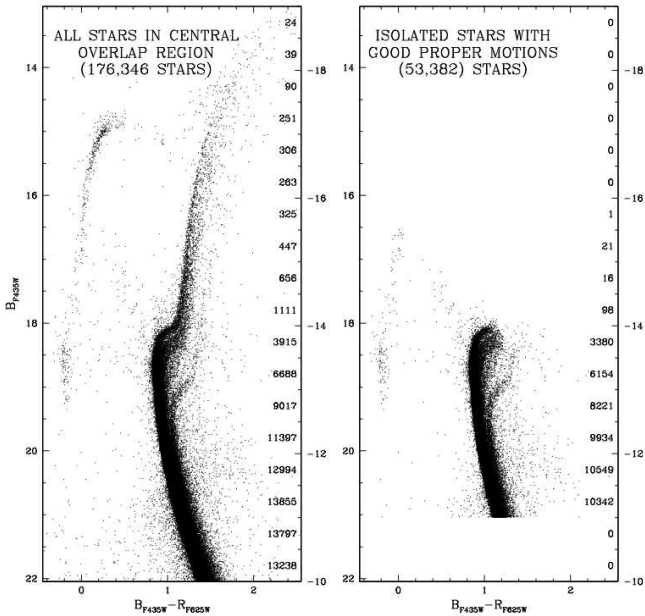}}
\figcaption{ Left:  all the stars from the catalog in the region of overlap
                 between the central pointing of GO-9442 and GO-10775.
         Right:  the stars found to have reliable proper motions.  
                 On the right of each CMD, we add up the number of
                 stars in each 1-magnitude bin.
         \label{fig09}}
\end{figure}
\fi

\subsubsection{The adjacent major-axis field}
\label{SSS.redux_PMs_otr}
In a similar way to that described above, we constructed proper motions for 
the adjacent field along the major axis.  Since the GO-10252 data straddled 
two of the GO-9442 pointings, we had to use two first-epoch pointings 
here, which doubled the number of exposures to handle, but the overall
reduction was generally the same.  To qualify for this proper-motion 
catalog, we required that a star be found in at least 4 first- and 4 
second-epoch exposures.  (Because of possible distortion errors at the
edges of the out-of-focus GO-10252 data set, we insisted on a star being 
found in at least 4 exposures to ensure that the internal errors will
be able to indicate when stars have distortion issues.)

We found 61,293 stars for which we could measure proper motions
in the adjacent field.  As above for the central field, we defined 
a high-quality sample of stars that should have uniformly good motions.
To qualify for this sample, a star had to have an RMS of measurements 
within each epoch of less than 0.02 pixel, and had to be brighter 
than $m_{\rm F435W} = -11$.  The high-quality proper-motion catalog 
for the adjacent field contains 19,593 stars.

\subsubsection{Proper Motion Zero-Points}
\label{SSS.redux_PMs_about}
The proper motions we have measured here are not absolute proper
motions, as is generally true for other globular clusters observed
with HST (see McLaughlin et al.\ 2006 for a broader discussion). HST
can measure absolute proper motions only when it has absolute
reference points to measure displacements against. The Omega Cen
fields are so crowded that there are no detectable galaxies that could
serve as absolute reference points. Also, the stability and
repeatability of HST (while superb) are not sufficient to compare
different frames directly. For example, there are small changes in
focus due to temperature-induced telescope breathing. These cause
scale changes that need to be calibrated out. Moreover, the nominal
orientation (roll) repeatability of HST is $0.003$ degrees. If left
uncalibrated, this would induce apparent solid-body rotation of the
cluster that exceeds the internal motions. Hence, to be able to derive
relative proper motions, we need to use general six-parameter linear
transformations to match exposures to each other.

The application of linear transformations implies that scale,
rotation, two skew terms, and two translation terms cannot be measured
from the data. The removal of scale and skew terms is not an issue,
because the dynamics of a cluster in equilibrium do not display
time-variability in these terms. The translation terms imply that we
can't measure the mean motion of Omega Cen with respect to the
Sun. This does not affect the internal cluster dynamics, which is our
primary concern here. Therefore, the only important fact to keep in
mind is that all proper motions reported here are measured modulo an
undetermined solid-body rotation component. We show in Section~5.2.3
of Paper~II that this does not impact our modeling. Ground-based
line-of-sight studies of Omega Cen reported by van de Ven et
al.~(2006) show that any true solid-body rotation in our central field
should be quite negligible, even though we can't measure it directly
here. Note in this context that any unaccounted for solid-body
rotation component would generally tend to {\it decrease} the IMBH
mass implied by models, since it would act to lower the kinematical
gradient between the inner and outer parts of the cluster.

To be able to remove small time-variations in high-order geometric
distortion terms, we have gone one step further in our calibration.
Namely, we measured each star relative to its local neighbors (see
Section~\ref{SSS.redux_PMs_cen}). All the neighbors used should be
moving with the cluster, but each will have some random internal
motion. We measured each star against hundreds of its neighbors, so
the random internal motions should average out. Therefore, the
proper-motion zeropoint is different for each star, and equals the
mean projected motion of cluster stars at its position on the sky. In
other words, all mean motion is removed, not just solid-body rotation.

The ``local-neighbor correction'' procedure allows the most accurate
measurement of the internal proper-motion dispersions for the cluster.
However, this correction is applied by choice, and not by necessity.
We would not want it to remove an important signal that is in fact
present in the data.  Mean motion in the radial direction should not
exist in an equilibrium cluster.  However, there could be differential
rotation (i.e., the part of the rotation curve that is not a linear
function of radius).  A Keplerian rise in rotation around an IMBH
would fall in this category.  There is no evidence for such
differential rotation in any existing data, but we can verify this
directly with our HST data.  Without the local-neighbor corrections,
the data are fully sensitive to small-scale internal rotations and
motions.

Figure~\ref{fig10} visualizes the size of the local-neighbor
corrections that were applied.  Our set of stars with good proper
motions were distilled into boxes 100 pixels on a side.  We computed
the proper motion for each star two ways.  First, we took the esimate
generated above, which involved removing from each position measured
in each image the average of the local neighbors.  Second, we skipped
this ``boresight'' correction, using only 6-parameter global linear
transformations to relate positions in each image to the master frame.
The difference between these two measured positions is shown as the
vector in the plot.  A vector that reaches the edge of the circle
corresponds to a displacement difference of 0.01 pixel.

\ifemulate
\begin{figure}[t]
\epsfxsize=0.99\hsize
\centerline{\epsfbox{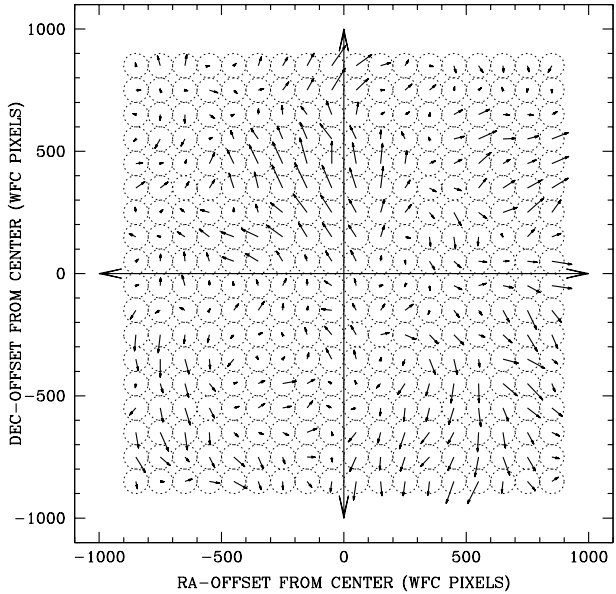}}
\figcaption{This figure exhibits the average local corrections applied
when deriving the proper motions, as function of position. The average
vectors were derived as described in the text for 18 $\times$ 18 bins
that cover the inner 1\minspt5 $\times$ 1\minspt5 . Vectors that
extend to the edge of the circle correspond to a displacement
difference of 0.01 pixel, which is typical of geometric distortion
residuals. The size of the corrections is too small (approximately one
eighth of the RMS cluster motion) to have any dynamical influence. The
morphology of the vector field indicates that we are not inadvertently
removing rotation intrinsic to Omega Cen.\label{fig10}}
\end{figure}
\fi

The vectors in Figure~\ref{fig10} correspond to the mean local motions
that were subtracted. Two things are evident. First, the corrections
are very small. The typical adjustment is less than 0.01 pixel, which
is characteristic for geometric-distortion residuals. This corresponds
to about an eighth of the observed proper-motion dispersion, and is
too small to have dynamical importance for the cluster. Second, there
is no overall rotation pattern visible. The correction field is patchy
and incoherent. Again, this is as expected for geometric-distortion
residuals, and this cannot be due to internal motions of the cluster.
More specifically, there is no excess rotation near the cluster center
that would be typical for Keplerian rotation around an IMBH.

As a further check, we also used the the methods of Paper~II to
determine the rotation curve (mean tangential proper motion as
function of radius) for the proper-motion catalogs with and without 
the local correction.  The difference in rotation curves was less 
than 1 km/s at all radii in the central field.

In summary, our catalog formally contains proper motions after
subtraction of all mean motion in the plane of the sky.  However,
detailed analysis and ground-based data indicate that any mean motion
that the cluster may in fact possess is dynamically irrelevant at the
central-field radii of our HST study.

In Sections~\ref{S.pmanal} and \ref{S.Exploits}, we present an initial
analysis of the proper motions.  Paper II will combine these motions
with existing proper-motion, radial-velocity, and surface-profile data
from the literature and will construct comprehensive dynamical models
for the cluster.

\subsection{Catalog and Image products}
\label{SS.redux_data_prod}
In the previous sections we described how the WFC images were analyzed 
to construct star and proper-motion catalogs.  These catalogs are
available along with the on-line version of this paper.

Specifically, we have analyzed the 10\arcmin\,$\times$\,10\arcmin\
field of dataset GO-9442, which goes out to beyond 2 core radii.  We
have one file ({\tt DATA.CATALOG.XYVI}) that contains 1,164,328 stars
found from running our automated finding algorithm on the F435W and
F675W images\footnote{
    Upon careful inspection of our bright stars, we found 112 bright 
    stars spread throughout the field in the initial catalog that were
    actually detector artifacts along the bleed-pixels of extremely 
    bright saturated stars.  We identified these stars easily as they
    had no counterpart at all in the H-alpha or the short images, thus
    we have removed them from our catalog.  This is the only 
    non-automated aspect of our catalog.}. 
Table~\ref{tab02} gives the column-by-column description
of this file. These measurements can be converted to calibrated
photometry using the zero points included in the table (from
Section~\ref{SS.redux_calib}).  We also provide the {\tt fits} images
referred to in Section~\ref{SS.redux_stack}, which contain a WCS
header and which are in the same coordinate system as the star
catalogs.


\begin{table}
\begin{center}
\caption{The columns in the star catalog, {\tt DATA.CATALOG.XYVI}.
         Columns 3 and 4 give instrumental magnitudes, which can be 
         transformed to the VEGAMAG scale by addition of the listed 
         zeropoints ZP$_{\rm VEGA}$.  Stars brighter than about $-13.75$ 
         are saturated in the deep exposures, and therefore have larger
         systematic errors.  We report the RMS scatter in individual 
         observations of each star.  In principle, the errors in columns 
         1 through 4 can be constructed by dividing these RMSs by the 
         square root of the number of observations (columns 10, 12 or
         their sum).  However, not all errors are random, so this may 
         underestimate the errors.  The RMS can be used to help
         select stars with good measurement consistency.
         }
\bigskip
\begin{tabular}{r|l}
\hline 
Column & Description                                                    \\
\hline 
1      & $x$ position in the reference frame                            \\
2      & $y$ position in the reference frame                            \\
3      & $m_{\rm F435W}$, instrumental $B$ magnitude, 
         ZP$_{\rm VEGA}$ = 32.043                                       \\
4      & $m_{\rm F625W}$, instrumental $R$ magnitude, 
         ZP$_{\rm VEGA}$  = 31.991                                      \\
5      & RMS scatter of the single-exposure $x$ positions               \\
6      & RMS scatter of the single-exposure $y$ positions               \\
7      & RMS scatter of the single-exposure $m_{\rm F435W}$ observations\\
8      & RMS scatter of the single-exposure $m_{\rm F625W}$ observations\\
9      & number of $B$ images where star could have been found          \\
10     & number of $B$ images where star was found well                 \\
11     & number of $R$ images where star could have been found          \\
12     & number of $R$ images where star was found well                 \\
\hline 
\end{tabular}
\label{tab02}
\end{center}
\end{table}

We also ran 500,000 artificial-star tests and subjected them to the same 
finding and measuring algorithm as was used on the real stars.  These 
tests are reported in file {\tt DATA.ART.XYVI}, which is also available 
in its entirety with the on-line article.  The artificial-star tests report 
the same measurement quantities as in Table~\ref{tab02} for the real 
stars; but for the AS tests, we also have the input positions as well,
recorded in {\tt DATA.ART.INPUT}.  It is worth pointing out that just 
because something was ``found'' in an artificial-star test, that does not 
mean that the input star was actually recovered; it is important to compare 
the input and output positions and fluxes to determine that the inserted 
star, and not a brighter pre-existing neighbor, was indeed recovered.

Finally, we provide the proper-motion measurements in separate files 
for the central and major-axis fields, {\tt DATA.PMs.CEN} and
{\tt DATA.PMs.MAJ}.  The central field has motions for 108,507 stars and
the adjacent major-axis field for 61,293 stars.  Table~\ref{tab03} gives a
column-by-column description of the proper-motion data files.  The 
``good'' flag refers to stars that are part of the high-quality subset
defined in Section~\ref{SSS.redux_PMs_cen}, which is used in the later
sections of this paper and in Paper II.


\begin{table}
\begin{center}
\caption{The columns in the proper-motion catalogs, {\tt DATA.PMs.CEN}
         and {\tt DATA.PMs.MAJ}.  The proper-motion errors are determined 
         by adding the errors in the average for each epoch in quadrature, 
         assuming no inter-epoch transformation error.  The major/minor 
         axis projections were made assuming the major axis to be 
         100$^{\circ}$ East of North.
         \bigskip}
\begin{tabular}{r|l}
\hline 
Column & Description                                                    \\
\hline 
   1  & $x$ position in the reference frame                            \\
   2  & $y$ position in the reference frame                            \\
   3  & $B_{\rm F435W}$, calibrated VEGAMAG $B$ magnitude \\
   4  & $R_{\rm F625W}$, calibrated VEGAMAG $R$ magnitude \\
   5  & $\mu_x$, proper motion in $x$, mas/yr                          \\
   6  & $\mu_y$, proper motion in $y$, mas/yr                          \\
   7  & $\sigma_{mu_x}$, proper-motion error in $x$, mas/yr            \\
   8  & $\sigma_{mu_y}$, proper-motion error in $y$, mas/yr            \\
   9  & $\mu_{\rm maj}$, proper motion along major axis                \\
  10  & $\mu_{\rm min}$, proper motion along minor axis                \\
  11  & $N_{\rm e1}$, number of exposures with good positions in epoch 1 \\
  12  & $N_{\rm e2}$, number of exposures with good positions in epoch 2 \\
  13  & $g$, the ``good'' flag, indicating small internal errors         \\
\hline 
\end{tabular}
\label{tab03}
\end{center}
\end{table}

In the following sections we will use these catalogs to determine the
center of the cluster (\S~\ref{S.center}), to extract a
surface-density profile (\S~\ref{S.sdp}), and to search for a cusp in
the velocity distribution that could be indicative of an IMBH at the
center (\S~\ref{S.pmanal}).  Finally, in \S~\ref{S.Exploits}, we will
compare the proper motions for stars in different populations.


\section{DETERMINING THE CENTER}
\label{S.center}
In general, the center of a cluster can be determined to a precision 
of about $\sigma / \sqrt{N}$, where $\sigma$ is roughly the core radius 
and describes the fall-off of the spatial distribution, and $N$ is the
number of stars used in the center determination.  For a cluster like 
Omega Cen, which has a core radius of 2\minspt5 (150\arcsec ), this 
means we would need over $150^2 = 25,000$ stars to measure a center 
to within an arcsecond.  These stars would have to be spread out 
to well beyond a core radius to give us a good handle on the center.
(The center is defined as much by where stars are {\it not} as by 
where they are.)

In the case of a Gaussian distribution, the most accurate way to measure 
the center is in fact to take a simple centroid of the star positions.  
However, there are several factors that can complicate this simple 
solution.  For example, if the region surveyed is not symmetric with 
respect to the center, then we must include some correction for this 
effect on the centroid.  Similarly, if the list of stars used in the 
determination is not complete throughout the field, then we must 
consider how this will affect the center determination.  The presence 
of bright giants can prevent us from finding the more plentiful 
main-sequence stars in their vicinity, and the small-number statistics
of the giants' distribution can result in patchy incompleteness.  All 
these limitations can be overcome; but it is important to address the 
issues carefully.

In this section we employ several different ways of determining
the center from our data. We determine the density center from star
counts, and the kinematic center from proper motions. In both cases we
use two separate methods to find the point of symmetry: one based on
contours and the other based on pie-slices. As an independent
cross-validation, we also derive the center of unresolved cluster light 
in a 2MASS mosaic image of the cluster. We find that all methods yield 
centers that agree to within the calculated errors.  We compare these 
centers with previous determinations available in the literature.
 
\subsection{Iso-Density Contour Centroids}
\label{SS.cen_cont}
The first way in which we determined the center was by constructing
iso-density contours, and fitting ellipses to them.  We began by 
sifting the stars in our 14000$\times$14000-pixel reference frame 
into an array of 140$\times$140 bins, where each bin is 100$\times$100
pixels (5\arcsec\,$\times$\,5\arcsec ).  We found that these bins 
did not contain enough stars to generate smooth contours, so we 
subsequently overbinned the distribution, making each bin correspond 
to 500$\times$500 pixels, or (25\arcsec\,$\times$\,25\arcsec ).  

The upper-left panel of Figure~\ref{fig11} shows the contours for 
the stars with $m_{\rm F435W} < -9$.  The edges of the field are clear, 
as is the region north of $(9000,8000)$ where there is a gap in the 
coverage (see Figure~\ref{fig01}.)  Our aim was to fit ellipses 
to the valid parts of these contours.  The contours traced in black are 
the ones that we trusted to be far enough from the edges of the field 
to be valid.

\ifemulate
\begin{figure}[t]
\epsfxsize=0.99\hsize
\centerline{\epsfbox{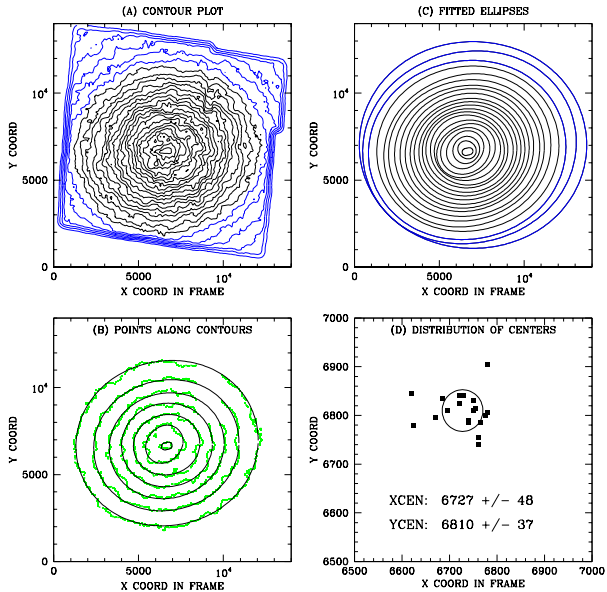}}
\figcaption{Panel A: The isodensity contours for all stars with $m_{F435W}<-9$.
                  Contours are separated by 100 stars (per 
                  25\arcsec\,$\times$\,25\arcsec\  box).  
                  The blue contours are contaminated by the edges of 
                  the field and are not used.
         Panel B: The points along contours 700, 1000, 1300, 1600, 1900,
                  2200, and 2500, and the ellipses that were fit to them.  
                  Points in the gap-region in the northwest were not used.  
         Panel C: The fitted ellipses (the parameters are given in 
                  Table~\ref{tab04}).
         Panel D: Distribution of the ellipse centers.  The circle is centered
                  on the average and has a radius equal to the RMS of
                  observations about the average. This RMS is also listed in 
                  the panel. The uncertainity in the average position is 
                  smaller
                  by $\sqrt{N}$, where $N \gtrsim 4$ is the number of 
                  {\it independent} contours.
         \label{fig11}}
\end{figure}
\fi

In panel (B) we show the points that fall along every third contour
(for clarity), and the ellipse that we fitted to each.  In these fits, 
along each contour we used only the points that were well away from the 
gap region.  Panel (C) shows the ellipses fit to all the contours.  The 
parameters of the fitted ellipses are given in Table~\ref{tab04}.


\begin{table}
\begin{center}
\caption{Summary of the fitted-ellipse parameters for the contours
         from stars brighter than $B_{435} = -9$.  The density is 
         reported in stars per 25\arcsec\,$\times$\,25\arcsec\  bin;
         the fitted center is pixels in the master frame; $A$ is 
         the semi-major axis and $B$ the semi-minor axis, in 
         WFC pixels; P.A. is the position angle of the major-axis, 
         measured in degrees E from N.  The last column represents 
         the ratio of the minor to the major axis. \bigskip }
\begin{tabular}{|r|c|c|r|r|}
\hline 
Density & Fitted center & ${{1}\over{2}} (B+A)$ & P.A. & $B/A$   \\
\hline 
  700   &  6780 6905  & 5055   & 100  &  0.87 \\
  800   &  6750 6810  & 4648   & 108  &  0.86 \\
  900   &  6695 6810  & 4320   & 104  &  0.86 \\
 1000   &  6740 6785  & 4018   & 102  &  0.86 \\
 1100   &  6720 6825  & 3726   & 106  &  0.87 \\
 1200   &  6740 6790  & 3471   & 106  &  0.88 \\
 1300   &  6760 6755  & 3207   & 104  &  0.88 \\
 1400   &  6765 6785  & 2982   & 102  &  0.89 \\
 1500   &  6780 6805  & 2749   & 102  &  0.89 \\
 1600   &  6775 6800  & 2539   &  96  &  0.89 \\
 1700   &  6720 6840  & 2303   &  92  &  0.90 \\
 1900   &  6755 6815  & 1841   & 102  &  0.87 \\
 2000   &  6670 6795  & 1618   & 118  &  0.80 \\
 2100   &  6625 6780  & 1387   & 118  &  0.86 \\
 2200   &  6620 6845  & 1082   & 134  &  0.83 \\
 2300   &  6685 6835  &  784   & 132  &  0.94 \\
 2400   &  6730 6840  &  526   &  56  &  0.87 \\
 2500   &  6760 6740  &  262   & 102  &  0.54 \\
\hline 
\end{tabular}
\label{tab04}
\end{center}
\end{table}

Panel (D) in Figure~\ref{fig11} shows the centers of the contours
shown in Panel (C).  The average position for the center is
(6727,6810) with an RMS of about 45 pixels in each coordinate.  Since
the contours are not entirely independent of each other, we cannot
simply construct an error in the center by dividing the RMS by the
square-root of the number of contours fit.  However, since we have at
least four independent contours (based on their separation and the
size of our bins), the center should be good to about 20 pixels, or
1\arcsec .

\subsection{The Pie-Slice Method}
\label{SS.cen_pie}
The focus of the previous approach was to find the center of the cluster
by identifying the centers of various iso-density contours of the stellar
distribution.  In this section, we describe how we divided the cluster 
into pie slices and determined the center by finding the location about
which the stars are most symmetrically distributed.  

The complication for this second approach is that the broad PSF halos
of the few bright giants can cause a patchy incompleteness in the
distribution of the faint stars, which can in turn cause us to
find more stars on one side of the cluster than on another, and thus
to misidentify the center.  The contour-based approach in the previous 
section skirted this issue somewhat by focusing on the outer contours,
where crowding is lower and the gradient is steeper; but the pie-slice
approach will need to deal with incompleteness all the way into the 
center. 

In order to measure an unbiased center, we had to construct a star list
that is free of any incompleteness that would bias the center.  This list 
must also contain enough stars to give us a statistically reliable 
result, so we could not simply use the few bright stars, which are
essentially 100\% complete.  We clearly need to include the more plentiful
faint stars, but including them requires taking careful account of 
incompleteness.  Our artificial-star tests from Section~\ref{SS.redux_AS} 
provided one way to assess the incompleteness in the field.  
Unfortunately, it was not feasible to run enough artificial-star 
tests to assess completeness on scales small enough to adequately 
sample the local region around each bright star.  What we 
needed was a simple map of the field that could tell us how bright a star 
must be in order to be found in a particular place in the field.  If we 
had such a map, we could then create a mask that is symmetric about any 
presumed center, so that the exclusion zones caused by the bright stars 
would be symmetric, and hence would not bias the center determination.

\subsubsection{Completeness Mask}
\label{SSS.cen_pie_mask}
Our goal in constructing a mask was to determine, for a particular bright
star, where in its vicinity a given faint star could be reliably found.  
To answer this question, we identified the bright stars with 
$m_{\rm F435W} < -16$ (about 10$\times$ saturation) and looked at the 
distribution of their found neighbors as a function of distance from 
them.  This distribution is shown in Figure~\ref{fig12}.  At a distance 
of 10 pixels, only stars that are within 5 magnitudes of the bright star's 
flux can be found; fainter stars than this cannot be found reliably.  Out to 
a distance of 20 pixels, stars up to 7.5 magnitudes fainter than the bright 
star can be reliably found.  This $d$ versus $\Delta m$ relationship is 
quantified by the red line drawn in the figure.  A few stars above this 
line are found, but essentially, {\it all} the stars below this line are 
found.

\ifemulate
\begin{figure}[t]
\epsfxsize=0.99\hsize
\centerline{\epsfbox{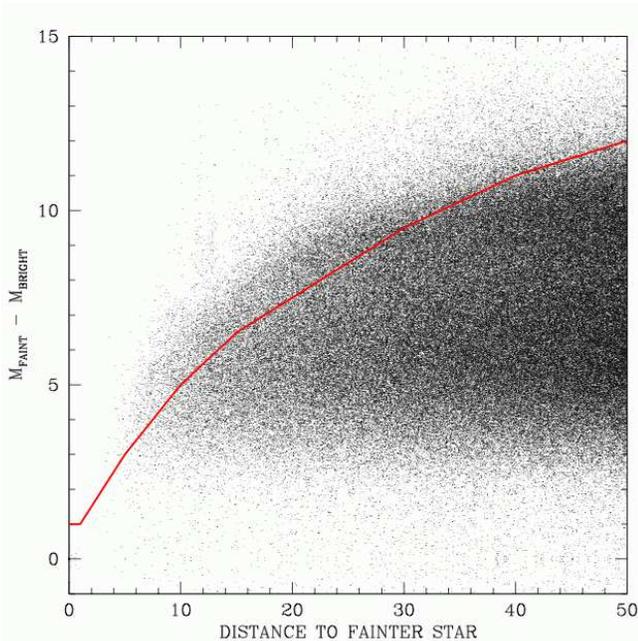}}
\figcaption{The distribution of found stars in the vicinity of brighter
         stars ($m_{F435W} < -16$).  The solid line (red in the on-line
         version) indicates the faintest magnitude of star that could 
         be definitively found at a given distance from a brighter star.
         \label{fig12}}
\end{figure}
\fi

Note that there are very few artifacts above this line; normally bumps
in the PSF at regular distances and offsets are identified as stars by
automated finding algorithms.  We have almost no such artifacts in our 
lists because our initial finding procedure had a model of the PSF out to
tens of pixels and the procedure was careful not to identify faint stars 
where appreciable PSF features are likely to be present (see Fig. 3 of A08). 
If we had identified every bump in every image as a star, then there 
would be a large number of detections in a distinct pattern above the 
line.  (Fig. 1 of McLaughlin et al.\ 2006 shows what the un-screened PSF 
artifacts look like in this space.) 

The red line in Figure~\ref{fig11} allows us to construct a zone around 
each bright star that tells us which stars stand out clearly at each 
distance.  We brought the information from all the bright stars 
across the entire field together in order to construct an image that 
tells us how bright a star must be at each point in the field in order
to be definitively found.  The value of each pixel in this image tells
us the magnitude of the faintest star that could be reliably found at 
each point.  Figure~\ref{fig13} shows the F435W image of the central 
region of the cluster.  The middle panel of the figure shows the 
finding mask constructed from these stars.

\ifemulate
\begin{figure}[t]
\epsfxsize=0.99\hsize
\centerline{\epsfbox{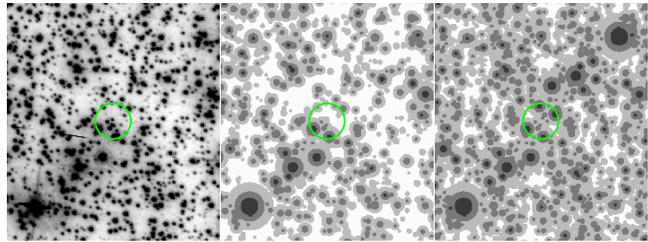}}
\figcaption{On the left we show a 10\arcsec\,$\times$\,10\arcsec\  F435W 
         image of the of central field.  In the middle panel, we show 
         the mask for the same region.  The mask shown here has been 
         gray-scaled to show the areas where stars of brightness $-8$, $-10$, 
         and $-12$ can and cannot be found.  The $m_{\rm F435W} \lesssim -8$ 
         stars can be reliably found only in the white areas.  The 
         $m_{\rm F435W} \lesssim -12$ stars can be found everywhere except for 
         the very darkest areas.  The actual mask used for the analysis 
         has much more gradation than is shown here.  On the right, we
         show the mask after it has been symmetrized about a presumed center
         (very close to the actual center).
         \label{fig13}}
\end{figure}
\fi

This mask essentially corresponds to an additional selection
requirement: stars must have been identified by our automated
procedure {\it and} they must also satisfy this location-based
requirement.  The goal of the mask is not to perfectly reflect our
artificial-star tests, but rather to provide an estimate of which
stars could definitively be found at various points in the field.  We
validated this assumption using the artificial-star tests: we verified
that artificial stars were indeed recovered essentially everywhere
the mask says they should be recovered.

One final note about the finding mask.  There is a gap in the north-west 
part of our field (see Fig~\ref{fig01}).  Since no stars could be found where
there was no coverage, we assigned the mask a value of $-25$ in this region.  
For reference, the brightest star in the field has an instrumental magnitude 
of $-20.28$ in F625W.  

\subsubsection{Symmetrization}
\label{SSS.cen_pie_symm}
The inferred mask made it possible to create star lists that have symmetric
incompleteness properties with respect to any adopted center.  With this 
tool in hand, we next determined the best center using a pie-slice procedure 
similar to that used in McLaughlin et al.\ (2006).  To do this, we assumed 
an array of trial centers.  About each trial center we divided the stars into 
pie slices centered upon that location and computed a statistic to compare 
the cumulative radial distributions of the stars in opposing pie slices.  
We finally identified the center as the place in the field that has the most 
similar radial distributions in the opposing pie slices.

Specifically, we decided to use eight pie slices, arrayed in cardinal 
and semi-cardinal directions (see Figure~\ref{fig14}).  We explored
centers within the range of reference-frame coordinates [6400:7200,6400:7200], 
with a trial center every 20 pixels (1\arcsec ) in each coordinate.  
In order to ensure that our star lists did not have any asymmetric 
biases, for each prospective center $(I_c,J_c)$, we generated a list 
of stars by asking whether each star in the catalog would be found 
both at its own location in the mask $(i_*,j_*)$, and at the location 
on the other side of the center $(I_c-[i_*\!-\!I_c],J_c-[j_*\!-\!J_c])$.  If 
the star satisfied both criteria, it qualified for the center-determination 
list.  The rightmost panel in Figure~\ref{fig13} shows what the mask 
looked like when symmetrized about a point near the ultimate center.  
A star that could not be found at one place in one slice would be excluded 
from the corresponding place in the opposing slice.

\ifemulate
\begin{figure}[t]
\epsfxsize=0.99\hsize
\centerline{\epsfbox{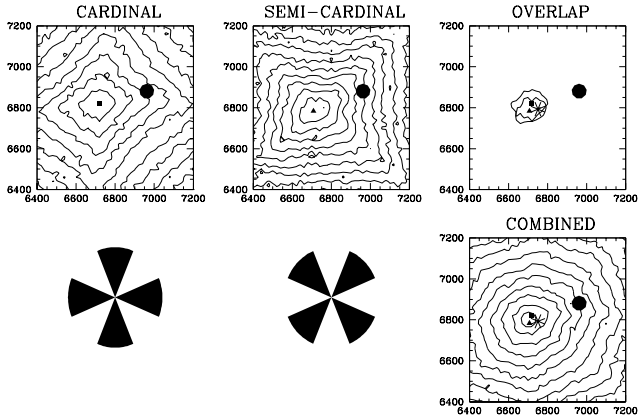}}
\figcaption{The two contour plots to the left show the goodness-of-center
         parameter for the pie slices shown below.  The symbol at the
         center represents the paraboloid fit to the 9$\times$9 points
         centered on each minimum.  In the upper right, we compare the
         innermost contours and the fitted centers.  In the lower right, 
         we add the two goodness-of-fit metrics to get an overall best
         fit. The filled circle is the center from NGB08, and the 
         6-pointed star is the center from the contour-based determination.
         The small square and triangle correspond to the center of the
         cardinal and semi-cardinal contours, respectively.
         \label{fig14}}
\end{figure}
\fi

We next selected the stars that both satisfied this symmetrized-mask 
criterion and were brighter than $m_{\rm F435W}\!=\!-9$  within 4000 pixels 
radial distance (200\arcsec ), and distributed them among the eight pie 
slices according to their azimuthal angle.  We sorted the stars in each 
slice by distance from the adopted center, then for each of the four opposing 
pie-slice pairs we computed $\sum_{n=min(N_1,N_2)} | r_1(n)-r_2(n)|$, 
which is the integrated difference between the radial distance $r$ versus 
cumulative $N_{\rm encl}$ distribution functions for the opposing slices.
This statistic was constructed for the four opposing pairs of slices at 
each trial center. 

Figure~\ref{fig14} shows the result for our pie-slice analysis.  The
four pairs of opposing slices allow us to construct two independent
estimates of the center, shown in the contours in the top plots.  We
fit a paraboloid to the central 9$\times$9 points in the contour plot
to arrive at the best-fit center for each panel.  The plot in the
upper-right panel shows the agreement between the two centers, and the
two central contours.  The plot on the bottom-right shows the contours
constructed from all 4 pairs of opposing slices.  The difference
between the two center estimates is 2\arcsec , indicating that our
average center is probably good to about 1\arcsec .  In this
determination of the center, we used about 235,000 stars.  With a core
radius of 150\arcsec , we would expect the center to be accurate to
about 150\arcsec\,$/\sqrt{235000}$ or about 0.3 arcsecond; this is not
a huge discrepancy if we consider that the cluster distribution is not
Gaussian and the star lists used here did not extend out beyond two 
core radii.

We conducted the same procedure with a subset of our stars, using only 
those stars with $m_{\rm F435W} < -10$ and $m_{\rm F435W} < -11$, and 
found very similar results.  Table~\ref{tab05} summarizes our findings.  
Overall, the centers determined by the different brightness cutoffs are 
consistent.


\begin{table}
\begin{center}
\caption{Summary of center determinations using the pie-slice method.  
         All positions are reported in our reference frame. \bigskip}
\begin{tabular}{|c|c||c|c||c|}
\hline 
Selection        & Number   & Cardinal    & Semi-cardinal & Together    \\
\hline 
$B_{435} < -9 $  & 235,000   & (6718,6820) & (6718,6820) & (6703,6801)  \\
$B_{435} < -10$  & 205,000   & (6723,6837) & (6738,6784) & (6723,6811)  \\
$B_{435} < -11$  & 150,000   & (6736,6854) & (6759,6807) & (6747,6829)  \\
\hline 
\end{tabular}
\label{tab05}
\end{center}
\end{table}
 
The center we found from the contour-based study in Section~\ref{SS.cen_cont}
was (6727,6810), which is in excellent agreement with the (6723,6811) 
location found here for $m_{\rm F435W} < -10$.  In what follows we will
adopt a center of (6725,6810) for the number density of Omega Cen, with
an estimated error of $\sim$1\arcsec\  in each coordinate.

\subsection{Kinematic Center from the Proper Motion Dispersion Field}
\label{SS.cen_kine}
An independent method to verify the results from the previous
subsections is to find the kinematic center. In the present context, that 
means identification of the point of symmetry in the proper-motion 
velocity dispersion field. Unlike the star-count analysis, the 
kinematic center determination has the advantage of being independent
of any incompleteness corrections (so long as the kinematics of a star
do not affect whether or not a star makes it into the catalog).

We took the full proper-motion catalog for the central field and
binned the stars onto a grid of $1'' \times 1''$ pixels. The few stars
believed to be outside the cluster (as indicated by
Figure~\ref{fig08}d) were excluded from the analysis.  For each pixel
we calculated the number of stars as well as the second proper-motion
moment ($\mu_x^2 + \mu_y^2$) summed over the stars. This yields two
images, which we will call $N_0(x,y)$ and $N_2(x,y)$. These images are
quite noisy, due to the relatively small number of stars per pixel. We
therefore applied a two-dimensional top-hat smoothing kernel to each
image to increase the signal-to-noise ratio. This yields images ${\hat
N}_0(x,y)$ and ${\hat N}_2(x,y)$. The implied smoothed image of the
one-dimensional RMS proper-motion $\sigma(x,y)$ is given by $[{\hat
N}_2 / (2{\hat N}_0)]^{1/2}(x,y)$. In practice we found that a top-hat
kernel radius of $25''$ yields adequate results.

The resulting RMS proper-motion image $\sigma(x,y)$ is shown in
Figure~\ref{fig15}. There is a well-defined symmetric distribution
around a broad central peak. The irregular outer boundary of the image
represents the intersection of the two (approximately square) ACS/WFC
fields for epochs 1 and~2, respectively. The cluster is not centered
within this boundary, due to the particular details of the
observational pointings. The closest boundary line is $\sim 70''$
South from the cluster center. In the kernel smoothing we excluded all
pixels that are outside the boundary region of the
catalog. Nonetheless, properties of the map within $25''$ from the
boundary may be somewhat affected by artifacts induced by the absence
of data outside the boundary. To ensure that this would not bias the
determination of the cluster center, we restricted our analysis of the
map to radii $R \lesssim 45''$ from the cluster center.

\ifemulate
\begin{figure}[t]
\epsfxsize=0.99\hsize
\centerline{\epsfbox{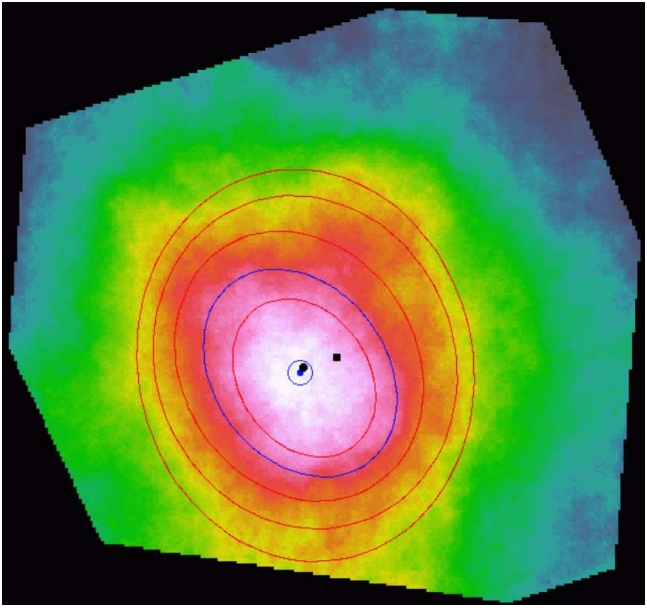}}
\figcaption{Velocity field of the one-dimensional proper-motion
dispersion, smoothed with a $25''$ radius top-hat kernel.  The
orientation of the image is as in Figure~\ref{fig11}.  The irregular
outer border of the image is the intersection of the two
(approximately square) ACS/WFC fields for epochs 1 and~2,
respectively. At the chosen level of smoothing, the gradient in
dispersion between the center and the edge of the field corresponds to
$\sim 2$ km/s. Ellipses are fits to the contours of the map, with
semi-major axis lengths of $30''$--$70''$. The lower dot (blue) is the
center of the ellipse (shown in blue) that has a semi-major axis
length of $40''$. This is used as an estimate of the HST kinematic
center, i.e., the symmetry point of the map. The small circle is the
68\% confidence contour around the kinematic center, as determined
from Monte-Carlo simulations. The confidence contour shows that the
kinematic center is consistent with the HST star count center (upper
dot, black). The center adopted by NGB08 (black square) is $12''$
away, and is inconsistent with both of the HST centers. 
\label{fig15}}
\end{figure}
\fi

We performed ellipse fits to the image to determine the symmetry point
of the map. We adopted the average ellipse center for semi-major axis
lengths between $35''$--$45''$ as our final estimate. The result lies
at $(\Delta_x,\Delta_y) = (-1.1'', -2.0'')$ from the adopted
star-count center. To determine the random errors in this center
determination we performed Monte-Carlo tests. We created pseudo data
sets by populating the same area of the sky covered by our catalog,
with the same number of stars. The stars were drawn from the projected
number density profile derived in Paper~II. Each star was assigned
proper motions in the $x$- and $y$-directions by drawing random
Gaussian deviates from distributions of dispersion $\sigma$. The value
of $\sigma$ was chosen to be a slowly decreasing function of radius,
consistent with our measurements in Section~\ref{S.pmanal} and
Paper~II. Proper motion errors were added based on Gaussian deviates
from randomly selected error bars in the observed catalog. Each pseudo
data set was analyzed in similar fashion as the real catalog. The
Monte-Carlo kinematic centers thus determined had an RMS scatter of
$2.9''$ per coordinate around the input center, with no bias. The
elongation of the $\sigma(x,y)$ contours in Figure~\ref{fig15} was
found not to be statistically significant. Ellipticities as large as
the observed value (at $R = 40''$) of $\sim 0.23$ happened by chance
in $14$\% of the Monte-Carlo simulations.

To further ensure the robustness of the results, we also experimented
with alternative approaches. These used percentiles of the
proper-motion distribution (instead of the RMS), different grid sizes,
different smoothing kernels, or different kernel sizes. The results
were always consistent with those quoted above. We also applied the
method to the high-quality proper motion subsample, instead of the
full proper motion catalog. This too yielded consistent results,
albeit with larger uncertainties.\footnote{Specific results were
included in an earlier preprint version of this paper. These are now
superseded by the analyses presented here.} This is because the full
catalog provides a larger sample size than the high-quality
subset. Although the full catalog has somewhat larger proper motion
uncertainties, these uncertainties have essentially no impact on the
analysis (although they are fully included in the simulations). They
are always much smaller than the cluster dispersion. Therefore, the
errors in the map of Figure~\ref{fig15} are determined primarily by
the number of stars that contribute to each pixel ($\Delta \sigma
\approx \sigma / \sqrt{2N}$), and not by the individual per-star
proper motion uncertainties. For the adopted kernel size, $N \approx
6000$ near the center. This yields random errors of $\sim 0.15$ km/s,
which is much smaller than the gradient in the dispersion map (in
fact, the size of the kernel was purposely chosen to make this the
case).

Our method is purposely designed to measure the symmetry point of the
proper motion map. One could use the peak of the map, but this
quantity is much more affected by shot noise, due to the intrinsically
low spatial gradients in the core of Omega Cen. Specifically, the peak
of the proper motion dispersion map in Figure~\ref{fig15} is at
$(\Delta_x,\Delta_y) = (-6'', -7'')$ from the adopted star-count
center. However, the Monte-Carlo simulations show that the peak pixel,
when used as an estimate of the cluster center, has an RMS uncertainty
of $6.7''$ per coordinate. This uncertainty gets even larger if one
chooses a smaller smoothing kernel size than the $25''$ used here. So
while the observed peak is statistically consistent with the symmetry
point, it is a much more unreliable estimate of the true cluster
center.

This makes an important point that underlies all our analyses: Omega
Cen has such a large and (nearly) homogeneous core, that to determine
its true center most accurately, one must locate the symmetry point
of its large-scale distribution. Measures of possible small-scale
peaks in density, star light, or kinematics, are much more susceptible
to shot noise. While this does not make such estimates incorrect, they
can only be interpreted if their uncertainties are rigorously
quantified.

\subsection{Kinematic Center from the Pie-Slice Method}
\label{SS.cen_kinpie}
As an alternative approach to determination of the kinematic center,
we also used a pie-slice method. For a given trial trial center, we
adopt a polar grid with $N$ azimuthal wedges ($N$ chosen to be even)
and $M$ radial bins along each wedge. The radial bins are linearly
spaced and have size $S$. The value of $M$ is chosen so that $M \times
S \approx 100''$, and the grid therefore encompasses most of our
central-field catalog. For each bin we use our full proper motion
catalog to calculate the proper motion dispersion $\sigma$ and its
error $\Delta \sigma$ as in Section~\ref{SS.cen_kine}. For each pair
of radially opposing bins we calculate the quantity $(\sigma_1 -
\sigma_2)^2 / (\Delta \sigma_1^2 + \Delta \sigma_2^2)$, which measures
the extent to which the dispersions in the two bins are statistically
consistent with each other. When summed over all the opposing bins,
this yields a $\chi^2$ quantity. The number of degrees of freedom
$N_{\rm DF}$ is normally $M \times N / 2$. However, we exclude pairs
that do not have a sufficient number of stars in each bin to yield a
meaningful dispersion. Each pair thus excluded reduces $N_{\rm DF}$ by
one. We map the quantity $\Delta \chi^2 \equiv \chi^2 - N_{\rm DF}$ on
a grid of trial centers with $1''$ spacing. We smooth the resulting
map with a Gaussian with a $2''$ dispersion (this smoothing is not
required, but was found to reduce the uncertainties in the resulting
kinematic center estimates by $\sim 15$\%). The trial center with the
lowest $\Delta \chi^2$ is the position with respect to which the
proper motion dispersion field is most radially symmetric.

To minimize any dependence of the results on the somewhat arbitrary
choices of $N$ and $M$, we repeated this procedure for
$N=4,6,8,10,12,14,16$ and $S=6,9,12,15,18,21,24$ arcsec. We summed the
$\Delta \chi^2$ maps from all the 49 combination of $N$ and $S$ to
obtain one grand-total $\Delta \chi^2$ map. The lowest $\Delta \chi^2$
in this map occurred at $(\Delta_x,\Delta_y) = (-0.4'', -1.5'')$ from
the adopted star-count center. We used the same pseudo-data sets as in
Section~\ref{SS.cen_kine} to characterize the uncertainties in the
result. Each pseudo data set was analyzed in similar fashion as the
real catalog. The Monte-Carlo results had an RMS scatter of $1.9''$ in
$x$ and $2.4''$ in $y$. This measures the uncertainties in the
inferred kinematic center. The Monte-Carlo results were found to be
slightly biased, and were centered on $(\Delta_x,\Delta_y) = (-0.4'',
-0.6'')$. Since this bias can be calculated and corrected, it is does
not affect the accuracy of the final result. The bias, and also the
fact that the scatters differ in the two coordinate directions, are
due to the specific geometry of our proper motion
dataset.\footnote{The bias can be avoided by using the method only
with data closer to the center, instead of going out to $100''$ where
the azimuthal coverage becomes incomplete. However, that would produce
larger uncertainties.}  Upon bias correction, the pie-slice estimate
for the kinematic center is $(\Delta_x,\Delta_y) = (0.0'' \pm 1.9'',
-0.9'' \pm 2.4'')$. We repeated the whole analysis also with only the
high-quality observed proper motion sample. This again yielded
consistent results, but with somewhat larger uncertainties.

The kinematic center thus determined using pie-slices is very
consistent with the value derived from contour fits in
Section~\ref{SS.cen_kine}. In fact, all methods that we have explored
to determine the kinematic center yielded the same answer to within the
uncertainties. The pie-slice method yields the lowest uncertainties,
because it uses more of the large-radius information in the catalog
(datapoints with $45'' \leq R \leq 100''$ are now included). We
therefore use the pie-slice result as our final estimate for the
kinematic center.

Our analysis provides the first time that the kinematic center of any
globular cluster has been accurately determined. The kinematic center
of Omega Cen was found to be consistent with the star-count
center. This is what is expected in an equilibrium system, and
therefore should not come as a surprise. However, this is important
since it provides an independent verification of the center position
determined from the star-count analysis.

\subsection{Comparison with the NGC08 Center and Other Literature Values}
\label{SS.cen_prev}
In \S\ref{SS.redux_refframe}, we used the 2MASS point-source catalog to tie 
our master reference frame to RA and Dec.  The center we identified in this
frame corresponds to $(\alpha,\delta)$ = (13:26:47.24, $-47$:28:46.45), 
with an error of about an arcsecond in each direction.

We were unable to find any other centers for Omega Cen in the
literature that had quoted errors.  The most recent determination of
the center was done by NGB08, who found the center to be at
(13:26:46.04,$-47$:28.44.8).  It turns out that this absolute position
was measured in the drizzled ACS images, and contains some systematic
error due to errors in the guide-star-catalog positions (Koekemoer et
al.\ 2005).  If we identify their center from the star field shown
in their Figure 2 (which corresponds closely to the absolute coordinate 
they report from the WCS header of image {\tt j6lp05weq\_drz}), then 
their center corresponds to (6962,6881) in our master frame, or
(13:26:46.08, $-47$:28:42.9) using the 2MASS astrometric reference frame, 
which is $(+237,+71)$ pixels or 12.3 arcseconds away from our center. 

Figure~\ref{fig18} provides in finding-chart format the locations of 
our star-count and kinematic centers determined from our HST analysis, 
with the error estimates indicated, as well as the center and IFU fields 
used by NGB08. NGB08 do not estimate the error in their center, so it 
is hard to say that our centers are formally in disagreement. However, 
given that they felt their center was within their 
5\arcsec\,$\times$\,5\arcsec\ IFU field, it seems safe to assume
our centers are in significant disagreement. We explore this further
in Section~\ref{SS.HSTcen_diff}.

\epsscale{1.00}
\ifemulate
\begin{figure}[t]
\epsfxsize=0.99\hsize
\centerline{\epsfbox{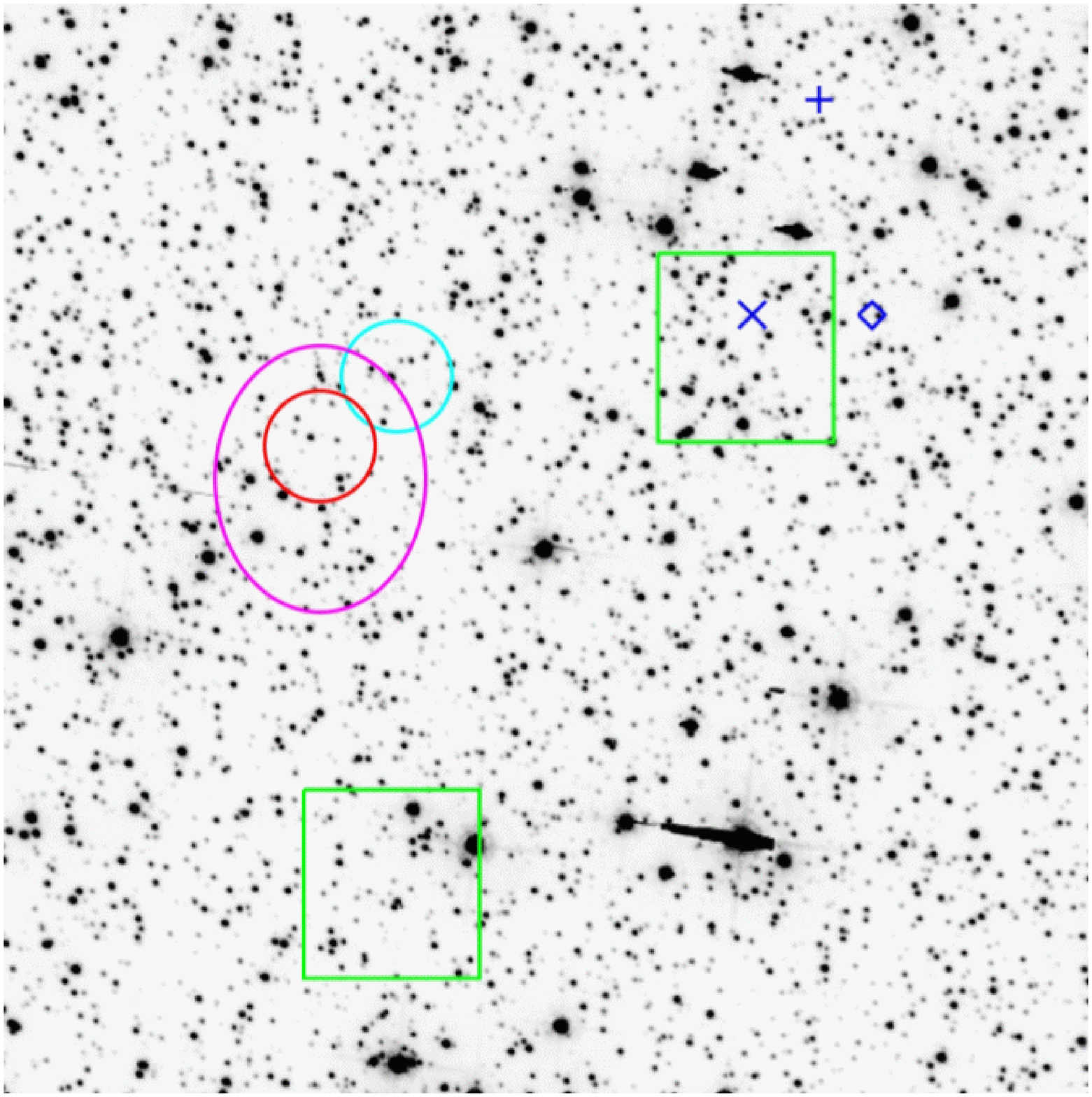}}
\figcaption{This figure shows the central 30\arcsec\,$\times$\,30\arcsec\
of Omega Cen. Estimates of the cluster center from Harris (1996;
plus), van Leeuwen et al.\ (2000; diamond), and NGB08 (cross) are
shown in blue. None of these authors provided error estimates for
their centers. The green boxes identify the two fields studied with
IFU spectroscopy by NGB08, one of which was believed to include the
cluster center. The circles/ellipses mark the cluster center positions
determined in the present paper using various independent methods,
namely HST star counts (red; Section~\ref{SS.cen_pie}), HST proper
motions (magenta; Section~\ref{SS.cen_kinpie}), and 2MASS unresolved
light (cyan; Section~\ref{SS.cen_2mass}). The sizes of the
circles/ellipses indicate the 68.3\% confidence regions of the
estimates (note: for a two-dimensional circular Gaussian probability
distribution, the circle that encloses 68.3\% of the probability has a
radius that equals $1.516$ times the one-dimensional error bar). Our
estimates are mutually consistent, and their uncertainties clearly
rule out the previously reported values in the literature (see
discussion in Sections~\ref{SS.cen_diff}
and~\ref{SS.HSTcen_diff}).\label{fig18}}
\end{figure}
\fi

Harris (1996) reports (13:26.45.9, $-47$:28:37) for the center in his
on-line catalog.  van Leeuwen et al.\ (2000) found a center of (13:26:45.756,
$-47$:28:42.780) using the positions of the giant stars brighter than
$B$ = 16. This center is adopted by van de Ven et al.\ (2006) and
Castellani et al.\ (2008). The Harris and van Leeuwen centers
are also indicated in Figure~\ref{fig18}. They are closer to the NGB08
center than to the center positions inferred by us. We explore the
differences between these ground-based centers and our HST center
further in Section~\ref{SS.cen_diff}.

In Table~\ref{tab06}, we summarize the centers from the literature and
compare them with the center determined here.

\subsection{Center from 2MASS}
\label{SS.cen_2mass}

Because of the disagreement between the center positions derived by us
and those previously published in the literature, we sought
non-HST-based data that could serve as an independent cross-check.
The publicly available data from 2MASS proved suitable for this
purpose.  We downloaded a wide 2MASS image mosaic of the cluster using
the NVO/ISRA mosaic service at
http://hachi.ipac.caltech.edu:8080/montage/ . The upper left panel of
Figure~\ref{fig16} shows the J-band mosaic image.  The pixel scale is
one arcsecond per pixel, and the field covers 14\arcmin\ $\times$
11\arcmin.

\ifemulate
\begin{figure}[t]
\epsfxsize=0.99\hsize
\centerline{\epsfbox{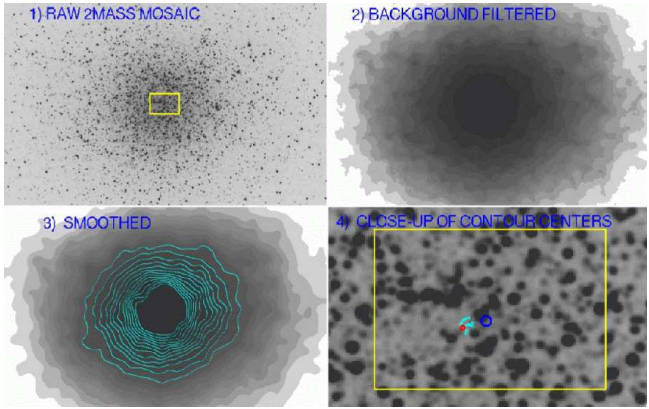}}
\figcaption{Upper left:  2MASS $J$-band mosaic centered on Omega Cen.
         Upper right: Result of filtering the image to derive the 
                      underlying background (as described in the text).
         Lower left:  Result after additional top-hat smoothing, with 
                      contours drawn in.
         Lower right: Close-up of the yellow-boxed region in the 
                      upper-left panel.  Locations of the contour centers 
                      are shown in cyan.  The red circle is the HST 
                      star-count center derived here and adopted  in our 
                      modeling; the 1\arcsec\ radius of the circle 
                      corresponds to the one-dimensional error bar. 
                      The NGB08 center is shown in blue.
         \label{fig16}}
\end{figure}
\fi

Since the 2MASS image was taken at near-IR wavelengths, it is
extremely sensitive to the red-giant stars. In order to minimize the
shot noise from bright stars on our analysis, we applied a ``filter''
to the image to isolate the underlying unresolved light. The filtering
operation consisted of going through the mosaic image pixel by pixel,
examining the surrounding pixels within a radius of 20 pixels
(20\arcsec ), and computing the tenth percentile of this neighbor
distribution.  The image on the top right of Figure~\ref{fig16} shows
the result.  It shows the general profile of the background light, but
it is still patchy on account of the few bright stars.

We next smoothed this background-filtered image with a circular
top-hat kernel that had a radius of 20 pixels, and arrived at the
image in the lower left panel of Figure~\ref{fig16}. This image is
smoother and allows the derivation of a contour-based center. The blue
lines are image contours between 2 core-radii and 0.75 core radii,
drawn at equal intervals of intensity. The adopted radial range is
well-suited for a determination of the center. It avoids the very
central region, where the spatial gradient is too low to yield high
accuracy. It also avoids, the outer region, where the results can be
biased by uncertainties in the 2MASS sky-background subtraction
process.

We fitted ellipses to the contours as in \S\ref{SS.cen_cont},
and show the resulting centers overplotted in cyan in the bottom right
panel of Figure~\ref{fig16}. This panel shows a close-up of the
yellow-boxed region in the upper-left panel. The average of the
centers lies at $(\Delta_x,\Delta_y) = (2.1'', 1.9'')$ from the
adopted HST star-count center. The RMS scatter in the ellipse centers
is $2.1''$ per coordinate. Given the kernel size for the
percentile-filter and blurring operations, we estimate that about a
third of the 12 contours between 2 core-radii and 0.75 core radii are
statistically independent. Therefore, the error in the mean position
per coordinate is $\sim 2.1'' / \sqrt{4} = 1.0''$. Given that the error
in the HST center is also $\sim 1''$, we find the centers determined
from HST and 2MASS to be in acceptable statistical agreement. By
contrast, the 2MASS position is $9.8''$ from the NGB08 center, and
even further from the Harris (1996) and van Leeuwen et al.\ (2000)
centers. It is therefore inconsistent with those centers.

Use of unresolved light, especially in the near-IR, is more
prone to possible systematic errors than our analyses based on HST
star counts and kinematics. Therefore, we would not assign the 2MASS
result the same level of confidence as our HST results, despite the
similar random error. However, the 2MASS result is useful as a
cross-check. The fact that it agrees with our HST results indicates
there is no reason to suspect some fundamental problem with using
either HST or ground-based data to determine the cluster center.

\subsection{Understanding the differences in ground-based centers}
\label{SS.cen_diff}

To understand how so many previous investigations could infer centers
that differ so much from the centers found here, we need to address
two separate questions: (1) why did previous authors who used
ground-based data, such as Harris (1996) and van Leeuwen et al.\
(2000), derive centers that differ so significantly from that derived
here from 2MASS?; and (2) why did NGB08, who used some of the same HST
data that we have, derive a center that differs so significantly from
our star-count and kinematic-center results? We address the first
question here. The second question is closely tied to the
determination of cluster number density profile from the HST data, and
we therefore discuss it in Section~\ref{SS.HSTcen_diff} below.

We simulated a ground-based image by combining the R-band photometry
from our (nearly complete) HST catalog with a broad, 3-arcsecond FWHM
PSF.  This is shown in the upper left panel of Figure~\ref{fig17}.  We
also generated an image that represents the number counts for stars
with $S/N > 100$ in the deep HST F625W exposures, distilled into the
same 0.5\arcsec-pixels as the simulated image. These medium-brightness
stars should not suffer much from incompleteness. The star-count image
is shown in the upper right panel of Figure~\ref{fig17}.

\ifemulate
\begin{figure}[t]
\epsfxsize=0.99\hsize
\centerline{\epsfbox{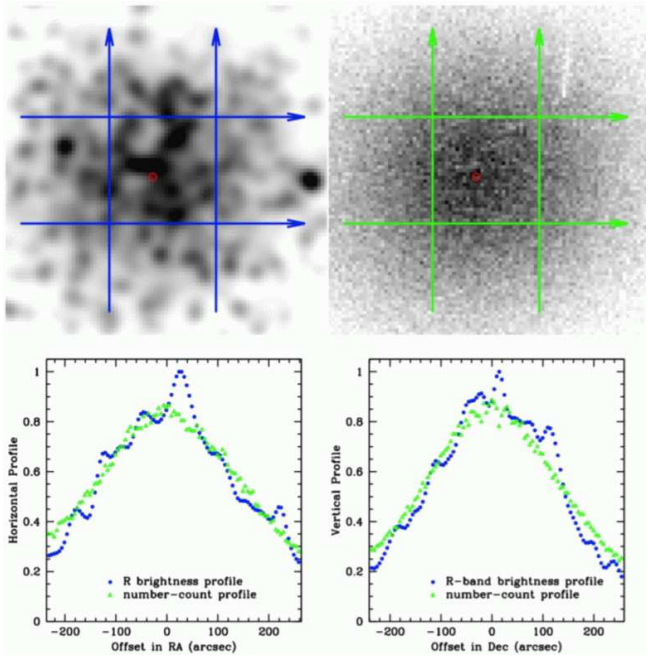}}
\figcaption{Upper left:  simulated image with 3\arcsec\ FWHM resolution 
                      of Omega Cen using our R catalog; 
         Upper right: number density of $S/N > 100$ stars from our catalog;
         Lower left:  horizontal profile of the two data sets;
         Lower right: vertical profile of the two data sets.
         The arrows delimit the regions used in the creation of the 
         profiles.  The red circle is the HST star-count center 
         derived here and adopted in our modeling; the 1\arcsec\ radius 
         of the circle corresponds to the one-dimensional error bar. 
         The ``traditional center'' (Harris 1996) adopted in many studies 
         of Omega Cen is at the center of the inner box.  The offset shown 
         on the horizontal axis of the bottom panels is measured with 
         respect to the HST star-count center.
         \label{fig17}}
\end{figure}
\fi

It is clear in the upper left panel that there are lines of bright 
stars in both dimensions that generate brightness enhancements
that cross at the rough location of the ``traditional center''. The
latter is at the center of the box, roughly 15\arcsec\ W and
10\arcsec\ N of the HST center derived above, which is marked by the
red circle. It is also clear that the visual centroid of the
bright-star distribution is N and W of the HST center that we have
identified.  At the same time, the upper right panel shows that the
bright, $S/N > 100$ sources are much more evenly distributed. They
appear centered, as expected, around the previously derived HST
center. Even by eye it is clearly evident that the $S/N > 100$ sources
are {\it not} centered on the center of the box. The dark areas of the
distribution extend more into the left and bottom parts of the panel
than they do into the right and top parts.

We extracted horizontal and vertical profiles from each of these
images. The lower left plot shows the horizontal profile across the
30\arcsec-tall swaths (between the horizontal arrows). The lower right
plot shows the same profile for the vertical direction. The blue
points correspond to the simulated R-band image, and the green points
to the number counts. It is clear that the traditional center
corresponds to a location where there is a coincidental concentration
of bright stars along both the $x$ and $y$ axes.

It is beyond the scope of this paper to determine how each
previous determination of the center went wrong. In fact, many of the
previous determinations did not provide a catalog or method
description of sufficient detail to allow easy reexamination. Also,
none of them provided an error estimate. Nevertheless, we have shown
that a few bright stars are responsible for a light enhancement that
is not coincident with the overall density peak of the more numerous,
but fainter, stars. This light enhancement is close to the positions
where previous authors estimated the cluster center to be. So it is
likely that the methods previously used on ground-based data were
disproportionately influenced by this light enhancement, and that this
led to the historical mis-identification of the center.

The analysis in Figure~\ref{fig17} does not imply that it is
impossible to derive the cluster center from ground-based data. After
all, we did manage to derive an accurate center from 2MASS
data. However, what is clear from these analyses is that at a minimum
one must adopt special methods to mitigate the shot noise from bright
stars (e.g., through filtering as used in our 2MASS analysis).


\section{SURFACE-DENSITY PROFILE}
\label{S.sdp}
Our nearly complete catalog of stars, covering the inner two core
radii of the cluster, allows us to calculate a definitive radial
density profile for the inner part of the cluster.  Note that the
entire cluster cannot be described by a single radial profile, since
the cluster contains stars of different masses at different stages of
relaxation.  If we consider only the profile of the evolved stars,
then we will suffer small-number statistics, since only about 1\% of
the stars in our catalog are above the SGB.  Our strategy here has
been to measure the number-density profile for stars within a range of
magnitudes.  We note that number-density profiles do a much better job
describing the star distribution than do surface-brightness profiles.
A star at the tip of the RGB is 100 times as bright as a star on the
SGB, yet the two stars have essentially the same mass and should both
be equally good tracers of the density distribution.  Since the
brighter stars are not better tracers of the star density, it makes
sense that we should not give them more weight.  Furthermore, since
there are many times fewer evolved stars than SGB and main-sequence
stars, surface-brightness profiles suffer much more from shot noise
than do number-density profiles.

\subsection{Profile from the new HST Catalog}
\label{SS.sdp_ourprof}

Our strategy in computing the number-density profile was straightforward.
We divided the field into concentric annuli centered on the center
derived in Section~\ref{S.center}.  Within each annulus, we determined 
the number of stars observed in each magnitude bin.  We also determined 
the completeness for the magnitude bin for that annulus, based on the 
artificial-star tests.  We then constructed an average surface density
within each annulus for each magnitude bin by dividing the number of 
found stars by the completeness.  

Figure~\ref{fig19} shows the surface-density profile for three
different brightness ranges, from the stars brighter than $m_{\rm F435W} =
-13$ (just above the turnoff), to stars just below the turnoff at
$m_{\rm F435W} \sim -11$, to stars well down the main sequence at 
$m_{\rm F435W} \sim -9$.  It is hard to know exactly which profile 
we are most interested in.  The profiles for the brightest stars tell 
us about the most massive luminous stars, but the fainter-star profiles 
have more stars, and therefore provide better constraints on the profile.  
All the profiles appear reasonably well fit by the single-mass King model
we provide as a reference.  Paper II fits the data with parameterized
models, which will serve as the basis for our dynamical modeling.

\ifemulate
\begin{figure}[t]
\epsfxsize=0.99\hsize
\centerline{\epsfbox{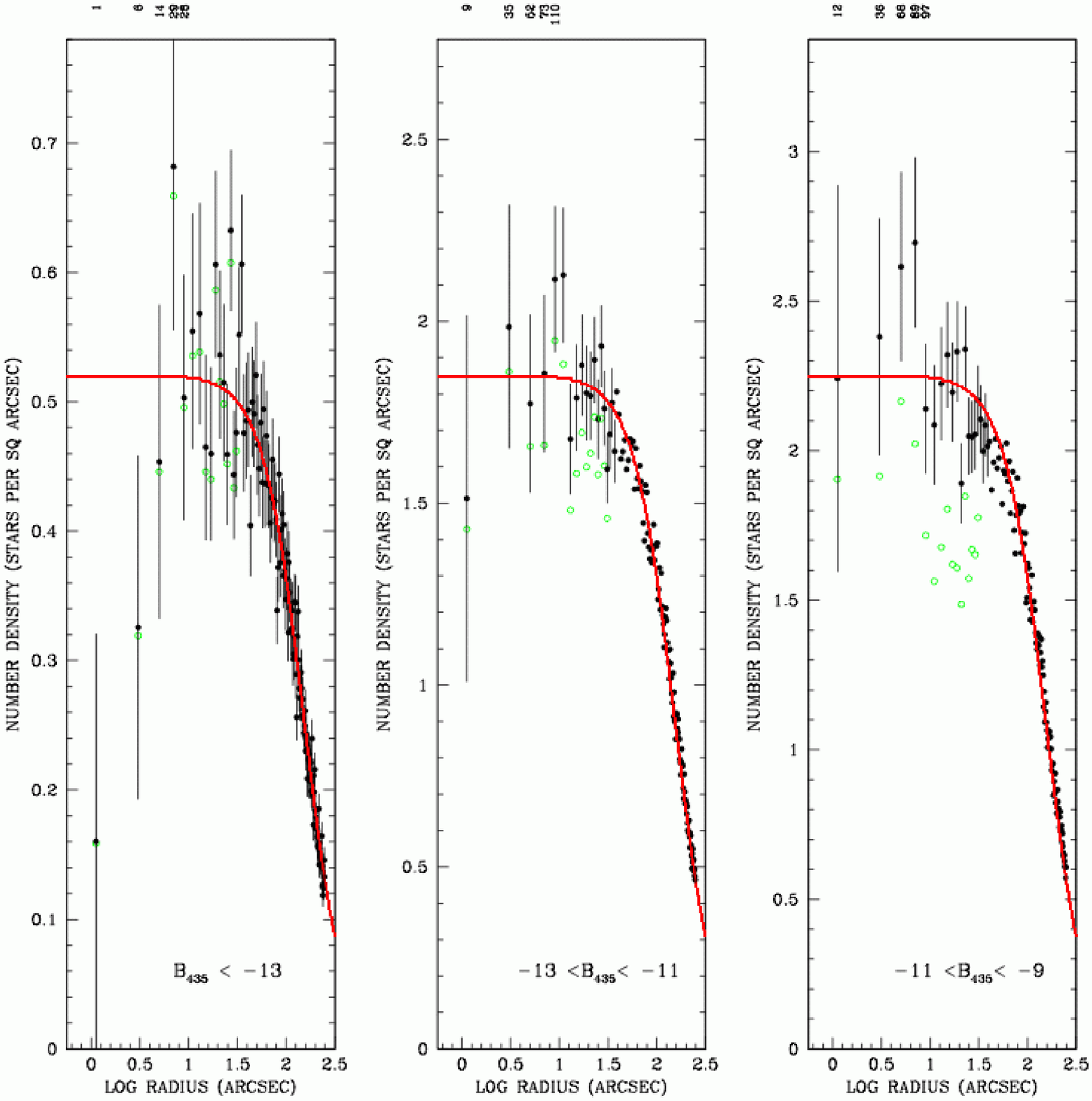}}
\figcaption{In each panel, we show the surface-density profile for 
         stars in the labeled magnitude range (in stars per square 
         arcsecond, with a linear scale).  For reference, the bottom 
         of the RGB is roughly at $m_{F435W} = -13.5$.  The filled 
         circles have been corrected for completeness.  The open 
         circles (green in the on-line edition) correspond to the 
         density of stars actually observed (for clarity, we plot 
         this only for the inner 50\arcsec ).  The error bars indicate
         the $\sqrt{N_{\rm obs}}$ errors.  The radial bins are 
         2 arcseconds wide; for the innermost 5 bins we note at 
         the top $N_{\rm obs}$, the actual number of stars observed.
         The curve (red in the on-line version) corresponds to a 
         single-mass King model with a core radius of 2\minspt5 
         and a tidal radius of 59\arcmin, shifted vertically 
         to fit each profile.  A more detailed quantitative analysis
         of model fits to the number density profile is presented
         in Paper II.
         \label{fig19}}
\end{figure}
\fi

There is no evidence of a sharp rise at the center in any of the
profiles.  Indeed, the central bin, which extends out to about
2\arcsec\ in radius, has fewer stars than the surrounding bins.
Section~\ref{S.center} showed that our center is accurate to about
1\arcsec , so the center should be well contained within our central
bin.  The completeness for all these bins is greater than 80\%, even
at the very center.  Looking at the image in Figure~\ref{fig02}, it is
clear that there is not a significant increase in the vicinity of the
center.  Quantitative constraints on the central slope of the number
density profile will be presented in Paper II.

\subsection{Comparison with the NGB08 Surface-Brightness Profile}
\label{SS.sdp_othprof}
In their paper, NGB08 measured the surface-{\it brightness} profile,
rather than the surface-{\it density} profile.  They found that it
rose with a power law with a logarithmic slope of $-0.08 \pm 0.03$,
which they deemed to be in significant disagreement with a flat core.
This is at odds with our result, in that we found a different center
and a relatively flat profile about it.

Since our data set is a super-set of the data they used, we decided to
try to reconstruct their profile using our star catalog.  In their
work, NGB08 measured bi-weights of the background light distribution
in the actual F435W image they used.  We approximated their
calculation by listing all the pixels within each of their annuli
(centered on their center) and taking the 25th percentile of the
points.  This is plotted as the filled black points in Panel (A) of
Figure~\ref{fig20}, and it agrees quite closely with the profile shown
in Figure~1 of NGB08.

\ifemulate
\begin{figure}[t]
\epsfxsize=0.99\hsize
\centerline{\epsfbox{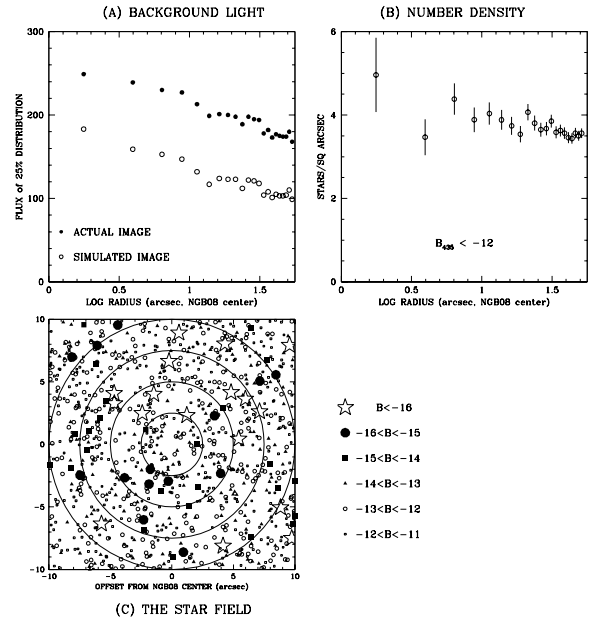}}
\figcaption{In Panel (a) we show that if we use the NGB08 radial bins, and 
         a percentile-based estimate of the background, we observe the
         same power-law radial trend they saw (filled black points).
         The open points represent the same procedure, but performed on a 
         simulated image, as described in the text, which contains no faint 
         unresolved stars.  Panel (b) shows the number-density profile
         about their center.  Panel (c) shows the catalog stars about the 
         NGB08 center.
         \label{fig20}}
\end{figure}
\fi

The NGB08 aim in measuring the surface brightness from the background
was to be less sensitive to the bright giants and more sensitive to
the numerous fainter stars.  Unfortunately, most of the light in the
background comes not from a large population of unresolved stars, but
rather from the PSF halos of the bright stars.  This can be seen from
a simple examination of the luminosity function.  In an external
galaxy where we cannot detect stars below the SGB, the luminosity
function (LF) is seen to increase steeply from the brightest stars to
well below the detection limit, and as such there is a large reservoir
of faint stars just below the detection limit. By contrast, in a
globular cluster such as Omega Cen, we resolve and count a large
fraction of all the stars that are present. The stars that are not
individually resolved contribute almost no light. In this sense, the
analysis of unresolved light from HST data is different than for
ground-based data. For example, the 2MASS images analyzed in
Section~\ref{SS.cen_2mass} do not resolve individual stars on or below
the SGB. So unlike the HST case, in the 2MASS images there is a large
repository of unresolved stars that contribute a significant amount of
light.

As a consequence of the fact that the LF is relatively flat below the
detection limit, most of the background in the HST images comes from
the bright-star halos rather than faint, undetected stars. Measuring
the profile of the HST image background is therefore equivalent to
measuring the profile from a blurred version of the bright stars.  To
demonstrate this, we simulated the field by taking the positions and
brightnesses of all the stars in our catalog and using PSFs that go
out to 100 pixels (5\arcsec ), based on the encircled-energy curves in
Sirianni et al.\ (2005).  We performed the same 25th-percentile-based
procedure on our simulated image, and arrived at the open dots, which
trace the actual observations extremely well.  There appears to be a
constant offset of $\sim$100 electrons between the real and the
simulated profiles, which can easily be accounted for by diffuse light
scattered by even more than 100 pixels---the bright core of this
cluster extends out to 3000 pixels (2.5 arcminutes).
 
In Panel (b) we show the {\it number-density} profile about the center
adopted by NGB08. This profile does not increase as steeply and
monotonically as the surface-brightness profile. To understand
qualitatively how the surface-brightness profile was found to be more
cuspy at the NGB08 center, Panel (c) shows the placement of the NGB08
apertures in the field.  Their central 50-pixel annulus happened to
just contain a large number of turnoff-brightness stars.  However,
there is no evident central concentration of the more plentiful
fainter stars, as a true cusp would suggest. The second annulus (from
50 to 100 pixels) contains a large number of bright RGB stars.  The
halos of these encircling stars may help to explain why the background
in the inner annulus was observed to be elevated.

\subsection{Understanding the differences in HST-based centers}
\label{SS.HSTcen_diff}

Both our study and the NGB08 study used HST data to determine the
center of Omega Cen. It is therefore of interest to examine in more
detail the differences between our results.

In Section~\ref{S.center} we have used the distribution of three
quantities to determine the center of Omega Cen, namely HST star
counts, HST proper motions, and unresolved 2MASS background light. For
each quantity we used a contour-based method, and for the HST-derived
quantities we also used a pie-slice method. The five different
analyses all give the same answer to within the uncertainties of 1--2
arcsec. This effectively rules out the possibility that unexplored
bias in any of the methods might have significantly affected the
analysis. What the methods all have in common is that they identify
the symmetry point of the cluster using data that extends
significantly from the cluster center. This naturally uses the full
size of the data set to reduce shot noise. The center thus identified
is $12.3''$ away from the NGB08 center. Hence, the NGB08 center is
ruled out as the symmetry point of the cluster at 5-10$\sigma$
confidence by each of 3 different quantities, analyzed with several
different methods.

NGB08 used a pie-slice approach on HST star count data to estimate the
cluster center. The general method they used is described in Noyola \&
Gebhardt (2006), but few details are provided about the specific
application to Omega Cen. Either way, there do appear to be two
important differences compared to our analysis in
Section~\ref{SS.cen_pie}. First, NGB08 applied no corrections for
incompleteness. We believe that this is generally inadvisable when
using star counts, although it is not clear whether this may have
specifically affected their analysis. Second, the NGB08 center was
measured from the same GO-9442 data set that we used, but from the
text of their paper it appears that they used only the central
pointing of the mosaic shown in Figure~\ref{fig01}a. This pointing was
centered on the traditional center of Omega Cen, which could
conceivably have introduced some bias. More importantly, the use of
only the central pointing implies that their radial coverage went out
to only 1\minspt5, which is significantly less than the core
radius. As a result, their analysis may have been more sensitive to
the peak of the density distribution rather than to the symmetry
point. This appears supported by the fact that their method did in
fact identify a density enhancement, as illustrated by
Figure~\ref{fig20}. This prompts the question whether there is any
physical significance to this enhancement.

The central part of Omega Cen has a large core that is almost
homogeneous. Within this core, spatial gradients in quantities of
interest tend to very shallow. When only a finite number of stars are
present or observed, one expects shot noise to dominate the
small-spatial density distribution within the core. Some areas may be
under-dense while others may be over-dense. This is in fact what we
appear to be seeing. At the position that we have identified as the
symmetry point of the cluster, the density appears somewhat underdense
compared to its surroundings (see Figure~\ref{fig19}). By contrast, at
the position that NGB08 identified as the cluster center, the density
appears somewhat overdense compared to its surroundings (see
Figure~\ref{fig20}). However, these features have the hallmarks of
being noise-induced valleys and peaks. As such, their presence has no
bearing on where the actual cluster center is.

The apparent underdensity at our center, as well as the apparent
overdensity at the NGB08 center, appear most prominent when the shot
noise is highest. In Figure~\ref{fig19}a this is because the number of
stars in the magnitude range $B_{435} < -13$ is low. In
Figure~\ref{fig20}a it is because of the use of unresolved light
(which emphasizes the shot noise from the brightest stars, which are
small in number). By contrast, in Figures~\ref{fig19}b,c and
Figure~\ref{fig20}b the respective under- and over-densities are less
prominent. Moreover, the errorbars are such that the profiles are
consistent with being flat to within the uncertainties. Specifically,
the rise in the number density within $10''$ from the NGB08 center
(Figure~\ref{fig20}b) is not statistically significant, and consistent
with being a statistical fluctuation. The error bar on the central
point is $\sim 20$\%, and the point just outside the center is lower
than the average within the inner $10"$.

Even a simple visual inspection of Figure~\ref{fig18} shows the core
of Omega Cen is sparse and homogeneous enough to make identification
of the center from local density enhancements either difficult or
impossible. Instead, one should adopt a technique that focuses on the
symmetry point of the larger-scale distribution. The data used for
such an analysis must extend far enough out to cover and use the
region where the density, brightness, and stellar motions start
dropping significantly. From such analyses we have found that the
NGB08 center is definitely not the symmetry point of the
cluster. Also, any light or density enhancement/cusp that may exist
near the NGB08 center appears consistent with a statistical
fluctuation in an otherwise (nearly) homogeneous core. On top of this,
we show in Section~\ref{SS.pmanal_cenfield} below that the proper
motions near the NGB08 center do not in any way indicate that this
position is special compared to its surroundings. We therefore
conclude that the position identified by NGB08 is not the cluster
center.


\section{ANALYZING THE PROPER MOTIONS}
\label{S.pmanal}
The best way to constrain the presence of an IMBH in a cluster is to
observe its effect on the motions of stars.  NGB08 used an IFU on
Gemini to measure the dispersion in the radial velocities of the
unresolved light, and inferred a distinct rise in the velocity
dispersion at their center, as compared with a field that was
14\arcsec\ away (see geometry in Figure~\ref{fig18}).  We have
demonstrated that the center they used is likely 12\arcsec\ off from
the true center.  Nonetheless, if their velocity measurements are
accurate, then they still imply an interesting kinematical feature
within the core.  To test this, we will examine the proper motions
about their center and about the center we derived.

\subsection{Proper Motions in the Central 10 arcsec}
\label{SS.pmanal_cen}

In Figure~\ref{fig21} we plot the total motions $\mu_D$ for the high-quality 
sample of stars (i.e., those flagged ``good'') in the
central ten arcseconds as a function of distance from the cluster
center. The total two-dimensional motions and errors are defined here
as
\begin{equation}
  \mu_D          = \sqrt{\mu_x^2 + \mu_y^2} , \qquad
  \sigma_{\mu_D} = \sqrt{\sigma_{\mu_x}^2 + \sigma_{\mu_y}^2} .
\end{equation}
The radial axis is scaled as $r^2$ so that we will get a roughly even
distribution of stars across the graph.  The vertical lines divide the 
sample into groups of 100 stars.  No star in the inner 10\arcsec\  is 
moving faster than 3 mas/yr.

\ifemulate
\begin{figure}[t]
\epsfxsize=0.99\hsize
\centerline{\epsfbox{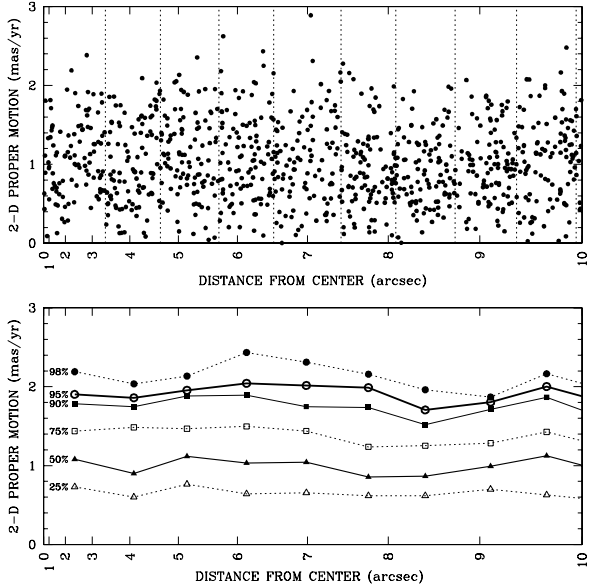}}
\figcaption{(Top) The total proper motion $\mu_D = \sqrt{\mu_x^2 + \mu_y^2}$ 
               for each star plotted as a function of distance from the 
               center. 
      (Bottom) The percentiles of the distribution for the 100-star
               bins indicated in the top panel. 
         \label{fig21}}
\end{figure}
\fi

This plot shows that the distribution of motions at the very center
(leftmost bin) looks very similar to that in the last bin at
$\sim$10\arcsec.  There are no more high-velocity stars at the very
center than elsewhere in the distribution, contrary to what we would
expect if an IMBH were present (Drukier \& Bailyn 2003).  The bottom
panel examines the percentiles of the bin-by-bin distributions.  In
addition to seeing no particularly fast-moving stars at the center, we
also see no indication of a rise in the dispersion at the center.
Consistent with our cursory examination above, the percentiles in the
central bin at $\sim$\,2 arcseconds are indistinguishable from those
in the outermost bin at 10 arcseconds.

Figure~\ref{fig21} indicates that the proper motions of Omega Cen 
do not show an obvious kinematical signature of an IMBH.  By
contrast, a sufficiently massive IMBH would have induced increasing
velocities towards the center with RMS $\sigma \propto R^{-1/2}$. The
question of what exact IMBH mass would be required to produce an
observable signature is discussed in detail in Paper~II. One important
issue when addressing this question is that many of the stars in the
centermost parts of the field are not truly close to the center but
are merely projected there from somewhere between $\sim \pm 1 $ core
radius along the line of sight. For example, for a projected aperture
of 3\arcsec\ radius around the center, only between 1.3\% and 5.8\% of
observed stars reside within 3\arcsec\ from the center in three
dimensions. These numbers were calculated from number density
distribution models derived in Paper~II (the ``core'' and ''cusp''
models, respectively). Only 43 stars in the high-quality subset of our
proper motion catalog reside within 3\arcsec\ from the projected
center. Hence, for an IMBH mass that produces a sphere of influence of
order {3\arcsec}, at most a handful of fast-moving stars would have
been expected (see the quantitative analysis in Section~6.7 of
Paper~II). The bigger the IMBH mass, the more fast-moving stars would
have been expected.

\subsection{Comparison with the NGB08 Kinematics}
\label{SS.pmanal_cenfield}

The analysis in the previous section examined the motions in the central
region out to 10 arcseconds from our center, but it did not include the 
central field studied in NGB08.  In Figure~\ref{fig22} we study the 
distribution of proper motions in the near vicinity of our center and 
the NGB08 center, in the context of the stars in the wider central region.
The top row of plots shows the proper motions for the stars in the inner 
15\arcsec; the middle and bottom rows show the same quantities for the
stars within 3\arcsec\ of our center and the NGC08 centers, respectively.
There are 1200 stars in the wider central region, 43 stars near our
center, and 54 stars near the NGB08 center.  We include here only stars 
from the high-quality proper-motion subset. 

\ifemulate
\begin{figure}[t]
\epsfxsize=0.99\hsize
\centerline{\epsfbox{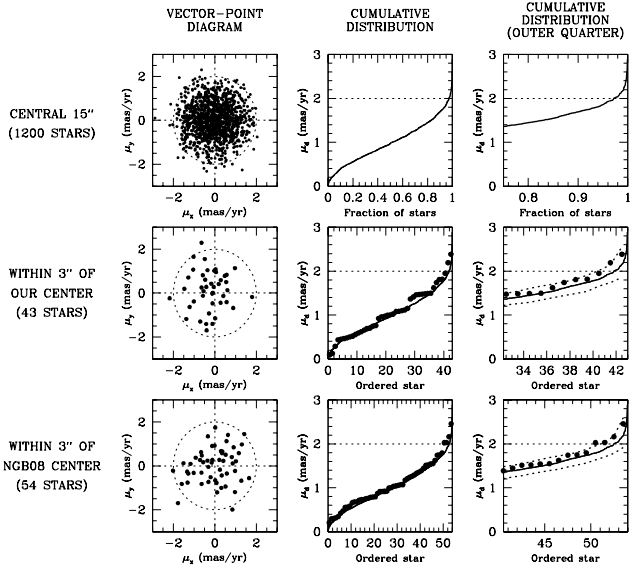}}
\figcaption{A comparison of the proper-motion distribution within the
         inner 15\arcsec\ (top row of panels) and the distribution within
         3\arcsec\ of the two centers (bottom rows).  The left panels
         show a vector-point diagram, the middle panels show the
         cumulative distribution of the two dimensional proper motions
         $\mu_D$, and the right panels show the outer 25\% of the
         cumulative distribution.  The dotted lines in the bottom
         right plot indicate the 80\% confidence region under the
         assumption that the underlying velocity distribution is the
         same as for the $ R \lesssim 15$\arcsec\ region in the top
         panel.  The observations fall within the bands, indicating
         that there are no statistically significant differences
         between the distributions.
         \label{fig22}}
\end{figure}
\fi

For each row, the leftmost panels show the proper-motion distribution 
as a vector-point diagram, and the right panels show the cumulative 
distributions.  For the bottom two rows, we compare the cumulative 
distribution for the given small region against the cumulative 
distribution for the central region as a whole.  The dotted lines in 
the right plots show the result of 1000 Monte Carlo tests based on 
the distribution function from the upper panel and the observed number
of stars in the lower two panels.  The dotted lines indicate the 80\%
confidence region under the assumption that the velocity distributions 
are the same.

We find that the motions of stars about our center and the NGB08
center are well described by the same distribution function that
describes the stars within the inner 15\arcsec ~of the cluster.
Therefore, our proper motions do not confirm the velocity gradient
reported by NGB08 in line-of-sight velocities, independent of the
adopted center.

To compare our kinematics more directly to those of NGB08, we also
studied the proper motions of stars in exactly the same two $5''
\times 5''$ fields that they studied (see Figure~\ref{fig18}). In our
high quality proper-motion sample, there are 51 and 35 stars in their
``central'' and ``off-center'' fields respectively. Each star has two
orthogonal velocity components, so the number of datapoints available
to estimate the one-dimensional velocity dispersion $\sigma_{1-D}$ is
twice the number of stars. Using the methods described in Paper~II, we
find that $\sigma_{1-D} = 0.833 \pm 0.059$ mas/yr and $0.835 \pm
0.071$ mas/yr, for the central and off-center fields,
respectively. For a canonical Omega Cen distance $D=4.8$ kpc (van de
Ven et al.~2006), these results translate to $18.9 \pm 1.3$ km/s and
$19.0 \pm 1.6$ km/s, respectively. The similarity between these
dispersions is consistent with the fact that we have determined both
to be at roughly the same distance $R=12''$ from the center (see
Figure~\ref{fig18}). These proper-motion dispersions are consistent
with the values we measure throughout the rest of the central $\sim
15''$ region of Omega Cen (see Paper~II). NGB08 determined
line-of-sight velocity dispersions $\sigma_{\rm los} = 23.0 \pm 2.0$
km/s and $\sigma_{\rm los} = 18.6 \pm 1.6$ km/s for these fields,
respectively.  Whereas our results agree with their measurement for
the off-center field, our proper motions do not confirm the NGB08
result that the velocities in central field are higher.

\subsection{Radial Proper-Motion Profile}
\label{SS.pmanal_rprof}
Figure~\ref{fig23} shows the proper-motion distribution for the entire 
central data set, defined by the overlap region between the GO-9442 and
GO-10775 data sets (see Fig.~\ref{fig01}), focusing on the percentiles
of the distribution function (as marked on the plot).  The plot on the
left is more finely sampled and goes out to a radius of 50 arcseconds.  
The plot on the right distills the stars into 2000-star bins and goes 
out to the corners of the central field, at 120\arcsec.  The radial axis
is once again scaled by $r^2$, so that we will get equal number of bins 
with radius (where there is full azimuthal coverage).

\ifemulate
\begin{figure}[t]
\epsfxsize=0.99\hsize
\centerline{\epsfbox{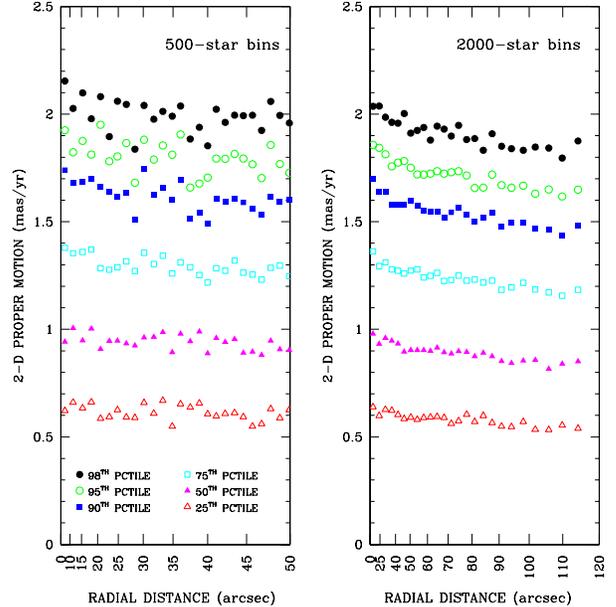}}
\figcaption{(Left) Similar to the bottom of Figure~\ref{fig21},
                but covering a larger radial extent.  We take
                the stars 500 at a time, and determine the percentiles of
                the proper-motion distribution, which we plot as a function
                of the median radius in the group.   
        (Right) Same, for 2000-star bins, out to the edge of the inner
                data set.  In a two-dimensional distribution, 50\% of 
                the points should be within 1.177 times the one-dimensional 
                sigma ($\sigma_{1-D}$), and 75\% should be 
                within 1.665 $\sigma_{1-D}$.
         \label{fig23}}
\end{figure}
\fi

While there is a distinct rise of about 10\% in the PM dispersion from 
80\arcsec\  into the center, the left plot shows that there is very little
increase within the inner 30\arcsec .  This is as true for the wings of the 
distribution function (the black symbols) as it is for the core (the cyan 
symbols).  

In Paper II we will present a detailed analysis of the proper motions
as a function of radius, including a determination of the
velocity-dispersion and velocity-anisotropy profiles, comparison to
literature data, calculation of higher-order Gauss-Hermite moments, a
comparison between the proper motions in our central and major-axis 
fields, and a comparison between major-axis and minor-axis proper motions.
Dynamical models will be fit to the data to constrain the possible
presence and mass of any IMBH.


\section{EXAMINING THE PROPER MOTIONS BY POPULATION}
\label{S.Exploits}
The primary motivation for the construction of a proper-motion catalog 
was to enable us to evaluate the likelihood of an intermediate-mass black 
hole at the cluster center (which we will further discuss in Paper II).
However, this rich data set also can tell us much about the dynamical 
state of the cluster.  In this section, we will do a cursory analysis 
of the data to examine the motions of stars with different masses and
stars in different populations.

\subsection{Equipartition}
\label{SS.equi}
It is well known that Omega Cen has not had enough time for complete
dynamical relaxation.  Harris (1996) reports its half-mass relaxation
time as $10^{10}$ years.  Anderson (2002) examined the luminosity
function at the cluster center and at a removed radius and found that
the cluster does not demonstrate as much mass segregation as one would
expect for a multi-mass King model in energy equipartition.  To
complicate the matter, Omega Cen has multiple populations of stars,
and these populations exhibit spatial gradients and could have
different mass functions. The present data set allows us to study the
dynamics of the stars at a single location in the cluster (the center),
thus avoiding the ambiguity introduced by comparison of populations at
different radii.

In Figure~\ref{fig24}, we show the distribution of the proper motions
and proper-motion errors as a function of F435W magnitude.  For this
analysis, we used the full PM catalog from the central field, not 
just the high-quality subset, since we wanted to examine the motions 
of the faint stars.  The color-magnitude diagram on the left highlights 
the stars that we selected for this study of the main sequence (MS).  
The middle panel shows the distribution of PM errors for the stars that 
lie within the curves in the left panel.  We selected for further analysis 
the stars that follow the general trend of PM error with magnitude in the 
middle panel, and show the motions for these stars on the right.  
Both $x$ and $y$ motions are shown in the same plot.  Within each 
half-magnitude bin, we found the error-corrected one-dimensional RMS, 
given the observed motions and errors for the stars in the bin.  
We report the RMS on the right-hand side of the plot.  The proper-motion 
dispersion clearly increases as we go down the main sequence.  The
RMS motions discussed here, as well as other kinematical quantities
discussed in the remainder of the paper, were determined using the
maximum-likelihood methodology described in Appendix A of Paper II.

\ifemulate
\begin{figure}[t]
\epsfxsize=0.99\hsize
\centerline{\epsfbox{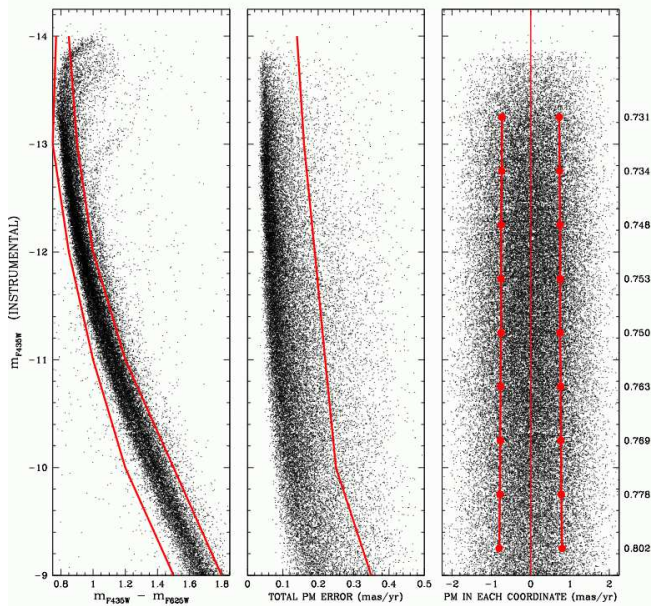}}
\figcaption{(Left) CMD showing stars selected for main-sequence study
                (those within the drawn-in lines).  
       (Middle) The two-dimensional proper-motion errors $\sigma_{\mu_D}$
                for stars as a function of instrumental F435W magnitude.  
                Stars left of the drawn-in curve were used in the analysis. 
       (Right)  The $x$ and $y$ proper motions for the selected stars
                plotted together.  The curves indicate the RMS as determined
                from a maximum-likelihood analysis that took into account the
                measured errors.  The single-component RMS for each 
                half-magnitude bin is reported on the right.
         \label{fig24}}
\end{figure}
\fi

In order to interpret this velocity variation in terms of mass, we fit
the CMD with an isochrone in the left panel of Figure~\ref{fig25},
finding a reasonable fit to the upper population with a 12.5 Gyr
isochrone from Pietrinferni et al.\ (2006) with alpha-enhancement,
[Fe/H] = $-1.6$ and Y=0.24.  This isochrone allows us to associate a
mass with each F435W magnitude.  The masses for our magnitude bins are
shown in Panel (b).  Finally, in Panel (c) we show (solid points) the
run of error-corrected RMS proper motion with mass.  If the cluster is
in energy equipartition, we would expect the observed points to follow
the upper curve, where velocity is proportional to $1/\sqrt{M}$.  We
see that the velocity does rise with decreasing mass, but not as
rapidly as equipartition would predict.  These results are consistent
with the core being in the process of establishing equipartition, but
only being about half-way there. Of course, it is well known that
Omega Cen cannot be represented by a single isochrone. Combined with
uncertainties in the exact distance, age, and metallicity, this causes
small uncertainties in the absolute calibration of the mass for a
given luminosity. However, this does not affect the conclusions about
equipartition, which depend only on relative masses.

\ifemulate
\begin{figure}[t]
\epsfxsize=0.99\hsize
\centerline{\epsfbox{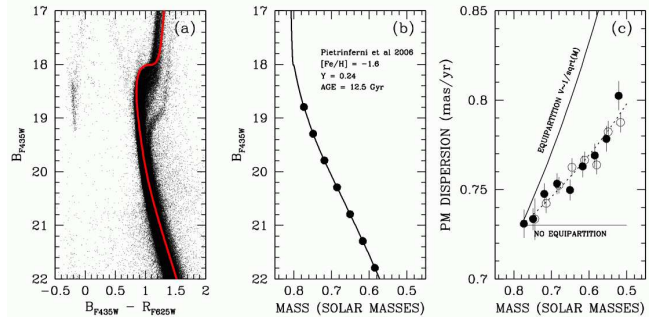}}
\figcaption{(Left) The fit of our CMD with the [Fe/H] $= -1.5$ and
                $Y=0.24$ isochrone from Pietrinfirni et al.~(2006).
       (Middle) The mass-luminosity relation for the isochrone, with 
                symbols at the locations of our luminosity-function 
                bins. 
        (Right) The observed error-corrected one-dimensional 
                proper-motion dispersion as a function of mass.  The 
                lines are drawn for comparison to indicate the trend 
                that would be expected for complete equipartition
                $\sigma \propto M^{-0.5}$ and for no equipartition
                at all.  The dotted line shows the intermediate case
                of $\sigma \propto M^{-0.2}$. Solid points show measurements
                for the central field; open points show measurements
                for the adjacent major-axis field, renormalized upward by
                a factor $1.22$.
                \label{fig25}} 
\end{figure}
\fi

We have repeated the central-field equipartition analysis also for the
adjacent major axis field. The median radius for stars in the
central-field catalog is $75''$, whereas it is $218''$ for the
major-axis field catalog. As a result of this difference in distance
from the cluster center, we find the RMS proper motion to be higher by
a factor of $1.22$ in the central field than in the major-axis
field. However, Figure~\ref{fig25}c shows that the dependence of RMS
proper motion on stellar mass is the same in both fields (open points
indicate the error-corrected RMS proper motion measurements for the
major axis-field after renormalization by a factor $1.22$). Therefore,
we detect no significant difference in the amount of equipartition
between approximately $0.5$ and $1.5$ core radii. In principle, one
might have expected less equipartition at the larger radius, given
that the two-body relaxation time increases with radius in Omega Cen
(see figure 21 of van de Ven et al.\ 2006). On the other hand, the
radial range that we can probe is not large. Also, cluster rotation
becomes a significant factor in the outer field, and this is not
probed by our analysis (as discussed in
Section~\ref{SSS.redux_PMs_about}). Detailed evolutionary models would
therefore be required for quantitative interpretation of these
results, which is outside of the scope of the present study.

In Paper II we will compare the observed dispersions of line-of-sight
velocities and proper motions in Omega Cen. The former are measured in
km/s, while the latter are measured in mas/yr. With the aid of
dynamical models, this provides a means of determining the cluster
distance.  However, mass segregation provides an added complication in
such an analysis. The stars for which we have proper motions are
slightly less massive than the stars for which we have radial
velocities (almost all are giant stars).  So even in the idealized
case of an isotropic system, one would not expect to measure
transverse and line-of-sight velocity dispersions that are the same in
km/s.  Figure~\ref{fig24} allows us to determine the size of this
effect and correct for it, as discussed in Paper II.

\subsection{Bulk motion of the metal-rich population}
\label{SS.smr_bulk}
Pancino et al.\ (2000) observed the RGB of Omega Cen from the ground 
in $B$ and $I$ and found evidence for three sub-populations:  a metal-poor
population (RGB-MP), an intermediate-metallicity population (RGB-MInt) and
a metal-rich population (RGB-a), which comprises 5\% of the RGB stars.  
Ferraro et al.\ (2002) then cross-identified these ``RGB-a'' stars in the 
proper-motion catalog of van Leeuwen et al.\ (2000) and found that they
appeared to be moving at about 1 mas/yr relative to the rest of the cluster.
They conjectured that this could represent a background cluster that may 
be in the process of merging with Omega Cen.  This interpretation 
was disputed by Platais et al.\ (2003), who contended that it was likely to 
be a consequence of an uncorrected color-magnitude term in the plate equation.
Recently, Bellini et al.\ (2009a) have constructed proper motions for stars 
in the outer regions of the cluster from ground-based CCD data spanning 
four years and determined that the RGB-a stars do in fact share the bulk 
motion of the cluster.  We will use our HST-measured proper motions and 
population-identifications to provide an additional determination of whether
the metal-rich population is moving with the cluster.  We focus here on the
lower-turnoff stars, the SGB analog of the RGB-a stars.

The left panel of Figure~\ref{fig26} shows a close up of turnoff region
for the stars for which we have good proper motions.  We highlight the
lower-turnoff (LTO) population in blue, and a control sample with similar 
flux in $R_{\rm F435W}$ (and thus similar astrometric errors) in green.
On the right, we show the proper motions for the two samples.  It is clear
that both distributions are centered on zero, meaning that they both share
the bulk motion of the cluster.  The cyan dot in the upper right panel 
shows the motion determined by Ferraro et al.\ (2002).  The center of 
the $\mu_x$ distribution for the LTO population is $-0.046 \pm 0.040$ mas/yr,
and the $\mu_y$ center is $-0.062 \pm 0.040$ mas/yr.  Both are consistent 
with zero, and similarly consistent with the motion for the control 
population.  We therefore conclude that all of the populations in the 
CMD are moving with the cluster and are phase-mixed.  We note that the
dispersions of the two populations are also the same, to within measurement
errors (3\%).

\ifemulate
\begin{figure}[t]
\epsfxsize=0.99\hsize
\centerline{\epsfbox{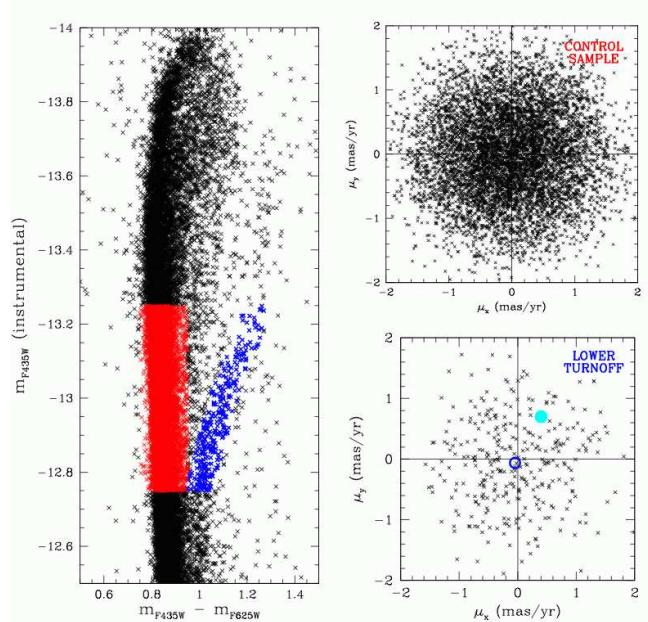}}
\figcaption{(Left) Color-magnitude diagram identifying the lower-turnoff
                population (blue in the on-line version) and the control 
                population (red), which should have similar 
                proper-motion errors.  
  (Lower right) The vector-point diagram for the control sample.  
  (Upper right) The vector-point diagram for the LTO population.  
                The circle represents the 90\%-confidence region
                for our fit to the center of the distribution.
                The filled dot (cyan) shows the motion as measured by 
                Ferraro et al.\ (2002).
         \label{fig26}}
\end{figure}
\fi

\subsection{Motions for the different MS populations}
\label{SS.bms_vs_rms}

Omega Cen was the first of the traditional globular clusters found
to have multiple populations.  The spread in metallicity along the giant
branch has been known for many decades (Dickens \& Woolley 1967 and
Freeman \& Rodgers 1975), but only recently have we been able to trace
the multiple populations down to the unevolved stars.  Anderson (2002) 
found and Bedin et al.\ (2004) confirmed that the main sequence clearly 
bifurcates into a red and blue branch below V $\sim$ 20.  Unexpectedly, 
they found that the fractions of stars in the two populations are opposite 
to what one would expect based on the RGB populations and standard enrichment
scenarios, which would have the bluer main-sequence population (bMS)
being more metal poor (and more populous) than the redder population 
(the rMS).  Norris (2004) pointed out that this could be explained by 
assuming that the metal-richer population was super-enriched in Helium.  
Villanova et al.\ (2007) then measured metallicities for stars in the 
two populations and found that the bluer stars were indeed more metal 
rich than the redder stars, in line with the Helium explanation.

Since these initial discoveries of the split MS populations, the
spatial distribution of the stars has been studied by Sollima et al.\
(2007) and by Bellini et al.\ (2009b).  They find that the
intermediate-metallicity population is more concentrated than the
metal-poor population, both when the populations are identified on the
RGB and on the main sequence.  Kinematically, it was initially
believed that the M-Int population did not share in the cluster
rotation (Norris \& Freeman 1997), but recent spectroscopy by Pancino
et al.\ (2007) mentioned above finds that all three populations appear
to share the same rotation, to within the 2 km/s measurement errors.

The radial velocities in the studies cited above come from all over the 
cluster, so they are able to probe the motions of stars with a global 
perspective.  Here, we have proper motions in only two fields, one at the 
cluster center and one at about a 1.5 core radius out, along the major axis.  
Furthermore, we have motions only for stars on or below the SGB, so it
is not possible to directly compare our motions against the radial 
velocities.  Nevertheless, we can still compare the motions of the 
populations we have access to.  In Figures~\ref{fig27} and \ref{fig28}, 
we compare the proper motions of the bMS and rMS stars in the central 
field and  the major-axis field, respectively.  

\ifemulate
\begin{figure}[t]
\epsfxsize=0.99\hsize
\centerline{\epsfbox{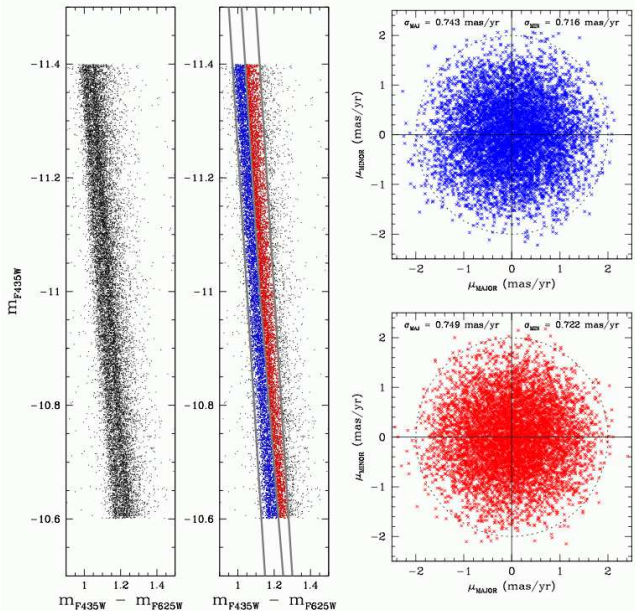}}
\figcaption{Dispersion for the main-sequence stars the central field.
         (Left) CMD before selection, 
       (Middle) CMD with selected bMS and rMS highlighted.  
  (Upper right) PM distribution of the bMS stars, with the 
                dispersions indicated at the top of the plot.
  (Lower right) Same for the rMS stars. Dashed circles are
                drawn for reference.  The error in each dispersion
                is about 0.006 mas/yr.
         \label{fig27}}
\end{figure}
\fi

Figure~\ref{fig27} shows motions for the bMS and rMS stars in the
central field.  The left panel shows the CMD in the location where the
two MS populations are clearly distinguishable in the central field, between 
an instrumental F435W magnitude of $-11.4$ and $-10.6$ (S/N $\sim$ 250).
The middle panel shows our selection of the bMS and rMS stars.  Finally, 
on the right, we show the vector-point diagrams with the motions for the 
two populations.  We analyzed the two distributions to determine the 
(error-corrected) dispersions as marked at the top of each plot.  The 
dispersions for the two populations in both the major- and minor-axis 
directions are the same to within the measurement errors (0.006 mas/yr).
For both MS populations, the major-axis dispersion is about 3\% greater 
than the minor-axis dispersion.  All of the distributions have no mean
motion, to within the measurement errors.

Figure~\ref{fig28} shows a similar plot for the adjacent field, which
is centered at a radius of about 4\arcmin\ ($\sim$1.5 $r_c$) along the
SE major axis (see Fig~\ref{fig01}).  Again, we identified the bMS and
rMS stars at a brightness along the MS where we could clearly
distinguish them and, at the same time where we had good proper
motions. The third column of panels shows the vector-point diagrams
for the two populations.  Again, the error-corrected dispersions noted
at the top show no statistically significant differences between the
populations. 

\ifemulate
\begin{figure}[t]
\epsfxsize=0.99\hsize
\centerline{\epsfbox{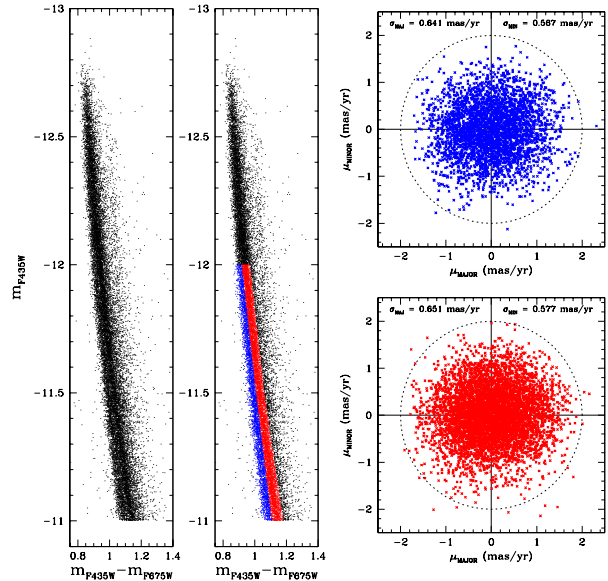}}
\figcaption{Similar to Figure~\ref{fig27}, but for the adjacent field.
         The error in each listed dispersion is $\sim$0.007 mas/yr.
         \label{fig28}}
\end{figure}
\fi

The dispersion on the major axis is larger than that on the minor
axis. This is what might naively have been expected from the fact that
the dispersion provides the pressure that supports the shape of the
system (rotation may generally contribute pressure as well, but we
show in Section~5.2.3 of Paper II this is negligible near the center
of the cluster). However, detailed axisymmetric anisotropic modeling
would be required to fully interpret the differences between major
axis and minor axis motion.

It bears repeating that any mean motion was removed from the proper 
motions during the data reduction stages, as discussed in 
Section~\ref{SSS.redux_PMs_about}.  Hence, the motions as we have 
measured them here are unable to measure rotation directly.  For that,
one would need reference objects to measure against (e.g., 
Anderson \& King 2003 measured the rotation of 47 Tuc in the plane 
of the sky using the background SMC stars). However, if there are 
multiple populations within the cluster and one population happens 
to be rotating relative to another, then we would expect this relative
rotation to manifest itself as a bulk motion between the two populations. 
This would be most prominent in the major-axis field, because
rotation in stellar systems typically decreases towards the center. 
Of course there would be no rotation in the plane of the sky if the
cluster were edge-on, but van de Ven et al.\ (2006) found the
inclination of Omega Cen to be 48$^\circ$.  Therefore we would expect
some difference in observed bulk motion if different populations had
different rotations.

Our analysis shows that the centers of the bMS and rMS distributions
are the same in the major-axis direction, but differ by 3-$\sigma$
(0.037 mas/yr, or 0.8 km/s, assuming a distance of 4.8 kpc) in the 
minor-axis direction.  This is consistent with the possibility that the 
populations are rotating relative to each other, but the statistical 
significance of the result is only marginal (especially when taking 
into account the possibility of small residual systematic effects at 
levels below 1 km/s).  We similarly examined the lower-turnoff population 
in the major-axis field and found that it shows no systematic motion 
relative to the other populations.  These findings are in agreement with 
the recent radial-velocity study by Pancino et al.\ (2007), who found 
that all the stars have the same rotational properties to within 2 km/s.

The motions we have measured here provide only an incomplete picture
into the dynamical state of the cluster.  Ideally, we would like to have 
both accurate proper motions and radial velocities, as well as 
population-identification information, for a large number of stars 
distributed throughout the cluster.  This would allow us to distinguish 
what kinds of orbits are populated by the different stellar populations, 
giving us much more information than simple studies of spatial 
distributions, dispersions, or rotational components.  To this end, 
we could augment our proper-motion catalog with motions for the brighter 
stars using the shorter exposures; there exist measured radial velocities 
for many of these stars and we would then have 5 of the 6 phase-space 
coordinates (lacking only the line-of-sight location).  However, the 
size of the central field would still be a significant limitation 
to our ability to infer the global properties of the cluster.


\section{Conclusions}
\label{S.concl}
We have performed a careful reduction of the large GO-9442 data set which
mosaic-imaged the inner 10\arcmin $\times$ 10\arcmin\ of Omega Cen with 
well-dithered HST/ACS observations. We constructed a 1.2-million-star 
photometric catalog of positions and $B_{\rm F435W}$ and $R_{\rm F625W}$ 
magnitudes from these data, along with a 14000$\times$14000-pixel stacked 
image of the field in the same reference frame as the catalog. We also
reduced the data from two other ACS/WFC programs that overlap with the
GO-9442 mosaic, but which covered only a single pointing (with medium-sized 
dithers).  GO-10775 observed a central field and GO-10252 an adjacent 
field to the southeast, roughly along the major axis. Accurate astrometric 
analysis of the data allowed us to determine proper motions for stars 
in these fields. The resulting proper-motion catalogs, with cuts applied 
to retain only the ``high quality'' measurements, contain 53,382 stars
in the central field and 19,593 stars in the major-axis field.  The data 
products from our study are made publicly available as part of this paper 
(see Section~\ref{SS.redux_data_prod}).

We analyzed the positions of the observed stars to determine the
cluster center. For this we used two separate methods, one based on
isodensity-contour fitting, and one based on the so-called
``pie-slice'' method. In the latter method we took particular care to
model the effects of incompleteness, and to correct for them. The
cluster centers thus determined each have an error bar of $\sim 1''$,
and they agree to within the errors. Upon use of stars in common with
the 2MASS catalog to calibrate to absolute coordinates, we find the
center to be at $(\alpha,\delta)$ = (13:26:47.24, $-47$:28:46.45). We
also used our proper-motion catalog to determine the kinematical
center of the cluster, defined as the symmetry point of the
proper-motion dispersion field on the projected plane of the sky. This
represents the first time that the kinematic center of a globular
cluster has been accurately determined. Again, we used methods based
on contours and pie-slices, with consistent results. The kinematical
center agrees with the star count center to within its $\sim 2''$
uncertainties. And finally, we also determined the center of
unresolved light in 2MASS data, again yielding a consistent result at
the $\sim 2''$ level.
 
We computed the surface number-density profile of the cluster around
its (newly determined) center. Artificial-star tests were used to
correct for the effects of photometric incompleteness. Density
profiles were determined for various ranges of stellar magnitude, but
were generally found to be similar, independent of magnitude. A
single-mass King model provides a reasonable fit to the inferred
profiles. There is no evidence for a strong number-density cusp
towards the center; in fact the density in the centermost bins appears
smaller (at marginal significance) than that at somewhat larger radii
($R \approx 10''$).

The proper-motion dispersion increases gently inwards from the core
radius ($\sim 2.5$ arcmin) to about $30''$, but flattens out at
smaller radii. Detailed analysis in the central $15''$ shows little
variation in kinematics with position. The dispersion does not
increase appreciably towards the center, and the wings of the 
proper-motion distribution do not become more extended towards the
center. There are no high-velocity outlier stars near the center that
might be indicative of motion around an IMBH.

We examined the variation of velocity dispersion with mass along the
main sequence and found that although the dispersion does increase for
the lighter stars, the cluster is not yet in equipartition. This is in
agreement with the findings in Anderson (2002), who found that the
cluster does not exhibit the mass segregation that would be expected
for a multi-mass King model in equipartition. These results are not
surprising, given the long half-mass relaxation time of $\sim 10^{9.96
\pm 0.03}$ years (McLaughlin \& van der Marel 2005).

Omega Cen has long been known to have multiple stellar populations.
The proper-motion catalog we have constructed here has enabled us to
look for variations in kinematics between populations. The blue (bMS)
and red (rMS) main sequence stars have very similar dynamical
properties. This holds for both the central field and the adjacent
field along the major axis, and is true both in terms of dispersion 
and mean rotation (the latter is not a trivial result, because 
Omega Cen is not believed to be edge-on; van de Ven et al.\ 2006).  The 
similarity in rotational properties contrasts with the findings of 
Norris \& Freeman (1997) based on radial velocities. However, it agrees
with a more recent study by Pancino et al. (2007). The mean motion of 
the metal-rich turn-off population is consistent with that of the rest
of the cluster. This is what would be expected for a (quasi-)equilibrium
configuration, and disagrees with an earlier finding of Ferraro et
al.\ (2002).

NGB08 argued for the presence of an IMBH in the center of Omega Cen
based on a combination of two arguments. First, they measured the
integrated line-of-sight velocity dispersion of unresolved light in
two $5'' \times 5''$ fields, one believed to be on the cluster center
and one at $R=14''$ from the center. The dispersion in the central
field ($23.0 \pm 2.0$ km/s) exceeded that in the off-center field
($18.6 \pm 1.6$ km/s), consistent with the presence of an IMBH of mass
$4.0_{-1.0}^{+0.75} \times 10^4 M_{\odot}$. Second, they measured the
surface brightness profile of unresolved light. They found it to have
a shallow central cusp of logarithmic slope $\gamma = 0.08 \pm 0.03$,
consistent with theoretical predictions for the cusp induced by an
IMBH (Baumgardt et al.\ 2005). The results from our new study have
allowed us to test these arguments.

We determined the one-dimensional proper-motion dispersion of the
stars in our catalog in each of the two fields studied by NGB08.  For
a canonical Omega Cen distance $D=4.8$ kpc (van de Ven et al.\ 2006),
the results translate to $18.9 \pm 1.3$ km/s and $19.0 \pm 1.6$ km/s,
for the central and off-center fields, respectively. So we find no
kinematical difference between the fields, and we also do not detect
kinematical gradients elsewhere in the central $\sim 15''$.  Moreover,
we find that NGB08 did not actually observe the cluster center in
their integral-field spectroscopy. The cluster center identified by
NGB08 is $12''$ from our newly determined center. We demonstrate that
this (and similar offsets in other previous determinations) is likely
due to biases induced by over-weighting of the small number of bright
giants, or the limited region over which stars were measured (the
central ACS chip in the case of NGB08, which covers only half the
core). Here we have included the more plentiful main sequence stars
and we have measured them out to beyond one core radius. This yields a
precise and unbiased handle on the center of the cluster density
distribution, as confirmed by our independent center determination
from HST proper motion and 2MASS unresolved light data.

The existence of a density cusp in Omega Cen, as reported by NGB08,
has also not been confirmed by our analysis. Since they calculated
their density profile around an incorrect estimate of the center, it
is unclear what physical meaning their result may have. Either way, we
showed here that use of unresolved integrated light, as was done by
NGB08, is not the best way to constrain the cluster density
profile. This method does not primarily measure the flux from a large
number of unresolved stars, but instead is sensitive to the
large-radii scattered PSF wings of bright giants. This method
therefore suffers more from shot noise (biases related to small-number
statistics) than a measurement of the number-density profile. This is
particularly important in a (nearly) homogeneous core such as that in
Omega Cen, since shot noise will always cause some areas of high
apparent density to exist by chance. The error bars on our number
density profile around the NGB08 center are such that the apparent
density enhancement there is consistent with being a statistical
fluctuation. The number density profile determined by us around the
newly determined center also shows little evidence of a significant
cusp towards the center, although a shallow cusp may not be ruled out.

In summary, our results do not validate the arguments put forward by
NGB08 to suspect the presence of an IMBH in Omega Cen. However, this
does not mean that such an IMBH may not be present after all. Our new
proper-motion catalog far exceeds the quality and quantity of the
kinematical data in the central arcmin previously available. This
provides the opportunity to study the central dynamics of Omega Cen
at a level of detail that is unmatched by almost all other clusters,
with the possible exception of 47 Tuc (McLaughlin et al.\ 2006). In
Paper II we therefore present a new detailed study of the dynamics and
density profile of Omega Cen, with the primary goal of exploiting the
new data to constrain the mass of any possible IMBH.

\acknowledgements

J.A. acknowledges the support from STScI grant GO-10401. We are
grateful to Ivan King and Mario Livio for feedback, advice and
encouragement during the course of this project.  Suggestions 
from the referees helped us improve the presentation of our 
results.



\begin{table}
\begin{center}
\caption{Various centers. (The NGB08 center has been corrected for HST 
         guidestar errors.) \bigskip}
\begin{tabular}{|c|ll|cc|ccc|}
\hline 
 Ref      & ~ ~ RA          & ~ ~ ~ Dec          & $x$  & $y$  &$\Delta x$ 
                                                       &$\Delta y$ 
                                                       &$\Delta r (\arcsec)$\\ 
\hline 
This work   & 13:26:47.24\  & $-47$:28:46.45 & 6725 & 6810 & ---  & --- &--- \\
NGB08       & 13:26:46.08\  & $-47$:28:42.9~ & 6962 & 6881 & +237 & +71 &12.3\\
Harris      & 13:26:45.9\ \ & $-47$:28:36.9~ & 7011 & 6960 & +286 &+150 &16.1\\
van Leeuwen & 13:26:45.756  & $-47$:28:42.78 & 7027 & 6883 & +302 & +73 &15.5\\ 
\hline 
\end{tabular}
\label{tab06}
\end{center}
\end{table}



\begin{references}

\parindent -0.10in
\narrower

Anderson, J.\ 2002, in ASP Conf. Ser. 265, 
    $\omega$ Centauri: A Unique Window into Astrophysics,
    ed. F. van Leeuwen, J. Hughes, \& G. Piotto (San Francisco: ASP), 87

Anderson, J., \& King, I.\ R.\ 2003, AJ, 126, 772

Anderson, J.\ 2005, in The 2005 HST Calibration Workshop, 
    ed. A.\ M. Koekemoer, P. Goudfrooij, \& L. Dressel (Baltimore: STScI)

Anderson, J., \& King, I.\ R. 2006, ACS/ISR 2006-01,
    PSFs, Photometry, and Astrometry for the ACS/WFC (Baltimore: STScI)


Anderson, J.\ et al.\ 2008, AJ, 135, 2055 (A08)

Baumardt, H., Makino, J., Hut, P., McMillan, S., \& 
    Portegies-Zwart, S.  2003a, ApJ, 589, L25

Baumgardt, H., Hut, P., Makino, J., McMillan, S., \& 
    Portegies-Zwart, S. 2003b, ApJ, 582, L21

Baumgardt, H., Makino, J., \& Hut, P. 2005, ApJ, 620, 238

Bedin, L.\ R., Piotto, G., Anderson, J., Cassisi, S., 
    King, I.\ R., Momany, Y., \& Carraro, G. 2004, ApJL, 605, 125	

Bellini, A., Piotto, G., Bedin, L.\ R., Anderson, J., Platais, I.,
    Momany, Y., Moretti, A.\ P., \& Ortolani, S. 2009a, AA, 493, 959

Bellini, A., Piotto, G., Bedin, L. R., King, I. R., Anderson, J., 
    Milone, A. P., Momany, Y. 2009b, A \& A, in press. 

Castellani, V.\ et al.\ 2007, ApJ, 663, 1021

Dickens, R.\  J., \& Woolley, R.\  v.\  d.\  R. 1967, R. Obs. Bull., 128, 255

Drukier, G.\  A., \& Bailyn, C. D. 2003, ApJL, 597, 125

Ferraro, F.\  R., Bellazzini, M., \& Pancino, E. 2002, ApJL, 573, 95


Freeman, K. C., \& Rodgers, A. W. 1975, ApJ, 201, L71

Fruchter, A. S., \& Hook, R. N. 2002, PASP, 114, 144

Gebhardt, K., Rich, R. M., \& Ho, L. C. 2002, ApJ, 578, L41

Gebhardt, K., Rich, R. M., \& Ho, L. C. 2005, ApJ, 634, 1093

Genzel, R. et al. 2003, ApJ, 594, 812

Gerssen, J., van der Marel, R. P., Gebhardt, K., Guhathakurta, P.,
    Peterson, R., Pryor, C. 2002, AJ, 124, 3270 (addendum: 2003, AJ, 125, 376)

Ghez, A. M. et al. 2005, ApJ, 620, 744

Harris, W. E. 1996, AJ, 112, 1487

Koekemoer, A.\  M., McLean, B., McMaster, M. \& Jenker, H. 2005,
    in The 2005 HST Calibration Workshop, p 417. A.M. Koekemoer,
    P. Goudfrooij, and L. L. Dressel, eds.  Baltimore, MD.

Kong, A. K. H. 2007, ApJ, 661, 875

Leonard, P. J. T., \& Merritt, D. 1989, ApJ, 339, 195


McLaughlin, D.\  E., Anderson, J., Meylan, G., Gebhardt, K., 
    Pryor, C., Minniti, D., \& Phinney, S. 2006, ApJS, 166, 249

McLaughlin, D.\  E., van der Marel, R. P. 2005, ApJS, 161, 304 

McNamara, B.\  J., Harrison, T.\  E. \& Anderson, J. 2003, ApJ, 595, 187

Norris, J.\  E., Freeman, K.\  C., Mayor, M., \& Seitzer, P.
    1997, ApJL, 487, 187

Norris, J.\  E. 2004, ApJL, 612, 25

Noyola, E., \& Gebhardt, K. 2006, AJ, 132, 447

Noyola, E., Gebhardt, K., \& Bergmann, M. 2008, ApJ, 676, 1008 (NGB08)

Pancino, E., Ferraro, F.\  R., Bellazzini, M., Piotto, G., 
    \& Zoccali, M. 2000, ApJL, 534, 83

Pancino, E., Galfo, A., Ferraro, F.\ R., \& Bellazzini, M. 2007, ApJL, 661, 155

Pietrinferni, A.,  Cassisi, S., Salaris, M., Castelli, F.  2006 ApJ, 642, 797

Platais, I., Wyse, R.\ G., Hebb, L., Lee, Y.-W., \& Rey, S.-C.
    2003, ApJL, 591, 127

Pooley, D., \& Rappaport, S. 2006, ApJ, 644, L45

Portegies Zwart, S. F., \& McMillan, S. L. W. 2002, ApJ, 576, 899

Sirianni, M.\  et al.\ 2005, PASP, 117, 1049

Skrutskie, M.\  F., et al.\ 2006, AJ, 131, 1163

Sollima, A., Ferraro, F.\  R., Bellazzini, M., Origlia, L., Straniero, O.
    \& Pancino, E. 2007, ApJ, 654, 915

Ulvestad, J. S., Greene, J. E., \& Ho, L. C. 2007, ApJ, 661, L151

van den Bosch, R., de Zeeuw, T., Gebhardt, K., Noyola, E., 
     \& van de Ven, Glenn.  2006, ApJ, 641, 852

van der Marel, R. P., Gerssen, J., Guhathakurta, P., Peterson, R. C., \&
    Gebhardt, K. 2002, AJ, 124, 3255

van der Marel, R. P. 2004, in Coevolution of Black Holes and Galaxies,
    L. C. Ho, ed., p. 37 (Cambridge: Cambridge University Press)

van der Marel, R. P.  \&  Anderson, Jay.  This volume.  (Paper II)

van Leeuwen, F., Le Poole, R. S., Reijns, R. A., 
    Freeman, K. C., \& de Zeeuw, P. T. 2000, A\&A, 360, 472

van de Ven, G., van den Bosch, R.\  C.\  E., Verolme, E.\  K., \& 
    de Zeeuw, P.\  T. 2006, A\&A, 445, 513

Villanova, S., Piotto, G., King, I. R., Anderson, J., 
   Bedin, L.\  R., Gratton, R.\  G., Cassisi, S., Momany, Y., 
   Bellini, A., Cool, A.\  M., Recio-Blanco, A.,
   \& Renzini, A. 2007, ApJ, 663, 296

\end{references}
\end{document}